\RequirePackage{lineno}
\documentclass[aps,prc,floatfix,superscriptaddress]{revtex4-1}
\usepackage{hyperref}

\usepackage{graphicx}
\usepackage{dcolumn}
\usepackage{amsmath}
\usepackage{supertabular}
\usepackage{multirow}
\RequirePackage{lineno}
\usepackage{color}
\usepackage{cancel}
\usepackage[normalem]{ulem}
\usepackage{array}
\usepackage{tabularx}
\usepackage[dvipsnames]{xcolor}
\def\bea {\begin{eqnarray}}
\def\eea {\end{eqnarray}}
\def\la {\langle}
\def\ra {\rangle}
\def\be {\begin{equation}}
\def\ee {\end{equation}}

\newcommand{\pT}{$p_{\rm{T}}$}
\newcommand{\sNN}{$\sqrt {{s_{\rm NN}}}$}
\newcommand{\Npart}{$\langle N_{{\rm part}}  \rangle$}
\newcommand{\gold}{Au+Au~}
\newcommand{\SigPK}{$\sigma^{1,1}_{p,k}$}
\newcommand{\SigQK}{$\sigma^{1,1}_{Q,k}$}
\newcommand{\SigQP}{$\sigma^{1,1}_{Q,p}$}
\newcommand{\SigpiK}{$\sigma^{1,1}_{\pi,k}$}
\newcommand{\SigpiP}{$\sigma^{1,1}_{\pi,p}$}

\begin{document}

\setlength\linenumbersep{0.1cm}
\title{ 
Erratum: Collision energy dependence of second-order off-diagonal and diagonal cumulants of net-charge, net-proton and net-kaon multiplicity distributions in Au+Au collisions [Phys. Rev. C 100, 014902 (2019)]} 



\affiliation{Abilene Christian University, Abilene, Texas   79699}
\affiliation{AGH University of Science and Technology, FPACS, Cracow 30-059, Poland}
\affiliation{Alikhanov Institute for Theoretical and Experimental Physics, Moscow 117218, Russia}
\affiliation{Argonne National Laboratory, Argonne, Illinois 60439}
\affiliation{Brookhaven National Laboratory, Upton, New York 11973}
\affiliation{University of California, Berkeley, California 94720}
\affiliation{University of California, Davis, California 95616}
\affiliation{University of California, Los Angeles, California 90095}
\affiliation{University of California, Riverside, California 92521}
\affiliation{Central China Normal University, Wuhan, Hubei 430079 }
\affiliation{University of Illinois at Chicago, Chicago, Illinois 60607}
\affiliation{Creighton University, Omaha, Nebraska 68178}
\affiliation{Czech Technical University in Prague, FNSPE, Prague 115 19, Czech Republic}
\affiliation{Technische Universit\"at Darmstadt, Darmstadt 64289, Germany}
\affiliation{E\"otv\"os Lor\'and University, Budapest, Hungary H-1117}
\affiliation{Frankfurt Institute for Advanced Studies FIAS, Frankfurt 60438, Germany}
\affiliation{Fudan University, Shanghai, 200433 }
\affiliation{University of Heidelberg, Heidelberg 69120, Germany }
\affiliation{University of Houston, Houston, Texas 77204}
\affiliation{Huzhou University, Huzhou, Zhejiang  313000}
\affiliation{Indiana University, Bloomington, Indiana 47408}
\affiliation{Institute of Physics, Bhubaneswar 751005, India}
\affiliation{University of Jammu, Jammu 180001, India}
\affiliation{Joint Institute for Nuclear Research, Dubna 141 980, Russia}
\affiliation{Kent State University, Kent, Ohio 44242}
\affiliation{University of Kentucky, Lexington, Kentucky 40506-0055}
\affiliation{Lawrence Berkeley National Laboratory, Berkeley, California 94720}
\affiliation{Lehigh University, Bethlehem, Pennsylvania 18015}
\affiliation{Max-Planck-Institut f\"ur Physik, Munich 80805, Germany}
\affiliation{Michigan State University, East Lansing, Michigan 48824}
\affiliation{National Research Nuclear University MEPhI, Moscow 115409, Russia}
\affiliation{National Institute of Science Education and Research, HBNI, Jatni 752050, India}
\affiliation{National Cheng Kung University, Tainan 70101 }
\affiliation{Nuclear Physics Institute of the CAS, Rez 250 68, Czech Republic}
\affiliation{Ohio State University, Columbus, Ohio 43210}
\affiliation{Institute of Nuclear Physics PAN, Cracow 31-342, Poland}
\affiliation{Panjab University, Chandigarh 160014, India}
\affiliation{Pennsylvania State University, University Park, Pennsylvania 16802}
\affiliation{NRC "Kurchatov Institute", Institute of High Energy Physics, Protvino 142281, Russia}
\affiliation{Purdue University, West Lafayette, Indiana 47907}
\affiliation{Pusan National University, Pusan 46241, Korea}
\affiliation{Rice University, Houston, Texas 77251}
\affiliation{Rutgers University, Piscataway, New Jersey 08854}
\affiliation{Universidade de S\~ao Paulo, S\~ao Paulo, Brazil 05314-970}
\affiliation{University of Science and Technology of China, Hefei, Anhui 230026}
\affiliation{Shandong University, Qingdao, Shandong 266237}
\affiliation{Shanghai Institute of Applied Physics, Chinese Academy of Sciences, Shanghai 201800}
\affiliation{Southern Connecticut State University, New Haven, Connecticut 06515}
\affiliation{State University of New York, Stony Brook, New York 11794}
\affiliation{Temple University, Philadelphia, Pennsylvania 19122}
\affiliation{Texas A\&M University, College Station, Texas 77843}
\affiliation{University of Texas, Austin, Texas 78712}
\affiliation{Tsinghua University, Beijing 100084}
\affiliation{University of Tsukuba, Tsukuba, Ibaraki 305-8571, Japan}
\affiliation{United States Naval Academy, Annapolis, Maryland 21402}
\affiliation{Valparaiso University, Valparaiso, Indiana 46383}
\affiliation{Variable Energy Cyclotron Centre, Kolkata 700064, India}
\affiliation{Warsaw University of Technology, Warsaw 00-661, Poland}
\affiliation{Wayne State University, Detroit, Michigan 48201}
\affiliation{Yale University, New Haven, Connecticut 06520}
\author{J.~Adam}\affiliation{Creighton University, Omaha, Nebraska 68178}
\author{L.~Adamczyk}\affiliation{AGH University of Science and Technology, FPACS, Cracow 30-059, Poland}
\author{J.~R.~Adams}\affiliation{Ohio State University, Columbus, Ohio 43210}
\author{J.~K.~Adkins}\affiliation{University of Kentucky, Lexington, Kentucky 40506-0055}
\author{G.~Agakishiev}\affiliation{Joint Institute for Nuclear Research, Dubna 141 980, Russia}
\author{M.~M.~Aggarwal}\affiliation{Panjab University, Chandigarh 160014, India}
\author{Z.~Ahammed}\affiliation{Variable Energy Cyclotron Centre, Kolkata 700064, India}
\author{I.~Alekseev}\affiliation{Alikhanov Institute for Theoretical and Experimental Physics, Moscow 117218, Russia}\affiliation{National Research Nuclear University MEPhI, Moscow 115409, Russia}
\author{D.~M.~Anderson}\affiliation{Texas A\&M University, College Station, Texas 77843}
\author{R.~Aoyama}\affiliation{University of Tsukuba, Tsukuba, Ibaraki 305-8571, Japan}
% ...

\collaboration{STAR Collaboration}\noaffiliation

\maketitle

In the original paper Phys. Rev. C 100, 014902 (2019), the first measurements of off-diagonal cumulants of net-charge, net-proton (a proxy for the net-baryon) and net-kaon (a proxy for the net-strangeness) were reported using the first phase of RHIC beam energy scan (BES-I) data~\cite{Adam:2019xmk}. 
The second-order mixed-cumulant ratios between net-proton and net-kaon ($C_{p,k} = \sigma^{1,1}_{p,k}/\sigma^{2}_{k}$) at different collision energies (\sNN = 7.7-200 GeV) show a good agreement with various model predictions. 
However, the mixed cumulants between net-charge and net-proton ($C_{Q,p} = \sigma^{1,1}_{Q,p}/\sigma^{2}_{p}$), as well as the mixed cumulants between net-charge and net-kaon ($C_{Q,k} = \sigma^{1,1}_{Q,k}/\sigma^{2}_{k}$), showed significant deviations from the model predictions. An increasing trend of these ratios as a function of collision energy in 0-5$\%$ central events was reported. Triggered by the theory papers~\cite{Vovchenko:2021xcs}, we realized that the excess correlations in \SigQP~and \SigQK~arise due to an artifact of assuming Q, p (or k) as mutually exclusive variables while correcting for the particle reconstruction efficiency effects. In this erratum, we address this issue. We now present the new observables $\sigma^{1,1}_{\pi,k}$, and $\sigma^{1,1}_{\pi,p}$ that avoid the above assumption. The previously observed increasing trend with energy in $\sigma^{1,1}_{Q,k}$, and $\sigma^{1,1}_{Q,p}$ is no longer seen in the new observables of $\sigma^{1,1}_{\pi,k}$, and $\sigma^{1,1}_{\pi,p}$.

%

In the original paper, the efficiency correction for $\sigma^{1,1}_{p,k}$, $\sigma^{1,1}_{Q,p}$ and $\sigma^{1,1}_{Q,k}$ is performed using the binomial efficiency correction method~\cite{Bzdak:2012ab,Bzdak:2013pha} assuming net-charge, net-proton, and net-kaon are mutually exclusive variables. In that case, the expression for the efficiency correction formula for \SigQP~is
 \begin{eqnarray}
  \sigma^{1,1}_{Q,p} (\text{Corrected}) &=& \frac{1}{\epsilon_{Q}\epsilon_{p}}\langle n_{Q}n_{p} \rangle - \frac{1}{\epsilon_{Q}}\langle n_{Q} \rangle \frac{1}{\epsilon_{p}}\langle n_{p} \rangle.
 \label{eqn:corr}
 \end{eqnarray}
  Here $\langle... \rangle$ represents average over events in a given centrality class. 
The $n_{Q}$ and $n_{p}$ are the measured net-charge and net-proton numbers within the acceptance of our measurement. The $\epsilon_{Q}$ and $\epsilon_{p}$ are the average efficiencies for inclusive charged particles and protons, respectively. A similar expression is also used for \SigQK. For inclusive charged particles no identification is performed -- only the charge state is measured using the STAR TPC by measuring its helix. But for estimation of efficiency $\epsilon_{Q}$, the weighted average of tracking efficiencies of protons, pions, and kaons are used.  
Recently, we discovered that the Eq.~\ref{eqn:corr} is not valid for the mutually inclusive variables like in $Q$-$p$ and $Q$-$k$ correlations. This is because inclusive charge particle multiplicity ($n_{Q}$) contains both protons ($n_{p}$) and kaons ($n_{k}$). This introduces a self-correlation in the previously considered efficiency correction procedure. A detailed discussion of this issue can be found in Refs.~\cite{Vovchenko:2021xcs,Chatterjee:2021hcr}.\\
To avoid this problem, we report the correlation between net-pion and net-proton ($\sigma^{1,1}_{\pi,p}$) and between net-pion and net-kaon ($\sigma^{1,1}_{\pi,k}$). This can help us address the problem of self-correlation in $C_{Q,p}$ and $C_{Q,k}$~\cite{Chatterjee:2021hcr}. The combination between net-proton and net-kaon was already published in Ref.~\cite{Adam:2019xmk}. 
Pions have been selected within $0.4<p_{T}<1.6$ GeV/$c$ using both TPC and TOF. To select pions a cut $|n\sigma_{\pi}|<2$ and $0.01<m^{2}<0.06$ (GeV/$c$)$^{2}$ has been applied. Proton and kaon identifications are same as the original paper~\cite{Adam:2019xmk}.
Using $\sigma^{1,1}_{\pi,p}$, $\sigma^{1,1}_{\pi,k}$, and $\sigma^{1,1}_{p,k}$ we can redefine the cumulant ratios $C_{Q,p}$ and $C_{Q,k}$ as follows:\\ 
\begin{eqnarray}
C_{Q^{\rm PID},p} = \frac{\sigma^{1,1}_{Q^{\rm PID},p}}{\sigma^{2}_{p}} = \frac{\sigma^{1,1}_{\pi,p}}{\sigma^{2}_{p}} + \frac{\sigma^{1,1}_{k,p}}{\sigma^{2}_{p}} + 1,
 \label{PIDQPratio}
\end{eqnarray}
\begin{eqnarray}
C_{Q^{\rm PID},k} = \frac{\sigma^{1,1}_{Q^{\rm PID},k}}{\sigma^{2}_{k}} = \frac{\sigma^{1,1}_{\pi,k}}{\sigma^{2}_{k}} + \frac{\sigma^{1,1}_{k,p}}{\sigma^{2}_{k}} + 1.
\label{PIDQKratio} 
\end{eqnarray}
Here $\sigma^{1,1}_{Q^{\rm PID},p} = \sigma^{1,1}_{\pi,p} + \sigma^{1,1}_{k,p} + \sigma^{2}_{p}$ and $\sigma^{1,1}_{Q^{\rm PID},k} = \sigma^{1,1}_{\pi,k} + \sigma^{1,1}_{p,k} + \sigma^{2}_{k}$.  The notation ``PID" is used to indicate that instead of using inclusive charged particles as in our original paper~\cite{Adam:2019xmk}, we are using a combination of identified pions, kaons, and protons.\\

\renewcommand\thefigure{\arabic{figure}}
\setcounter{figure}{1} 
\begin{figure*}[th]
 	\centering
  	 \includegraphics[width=\textwidth]{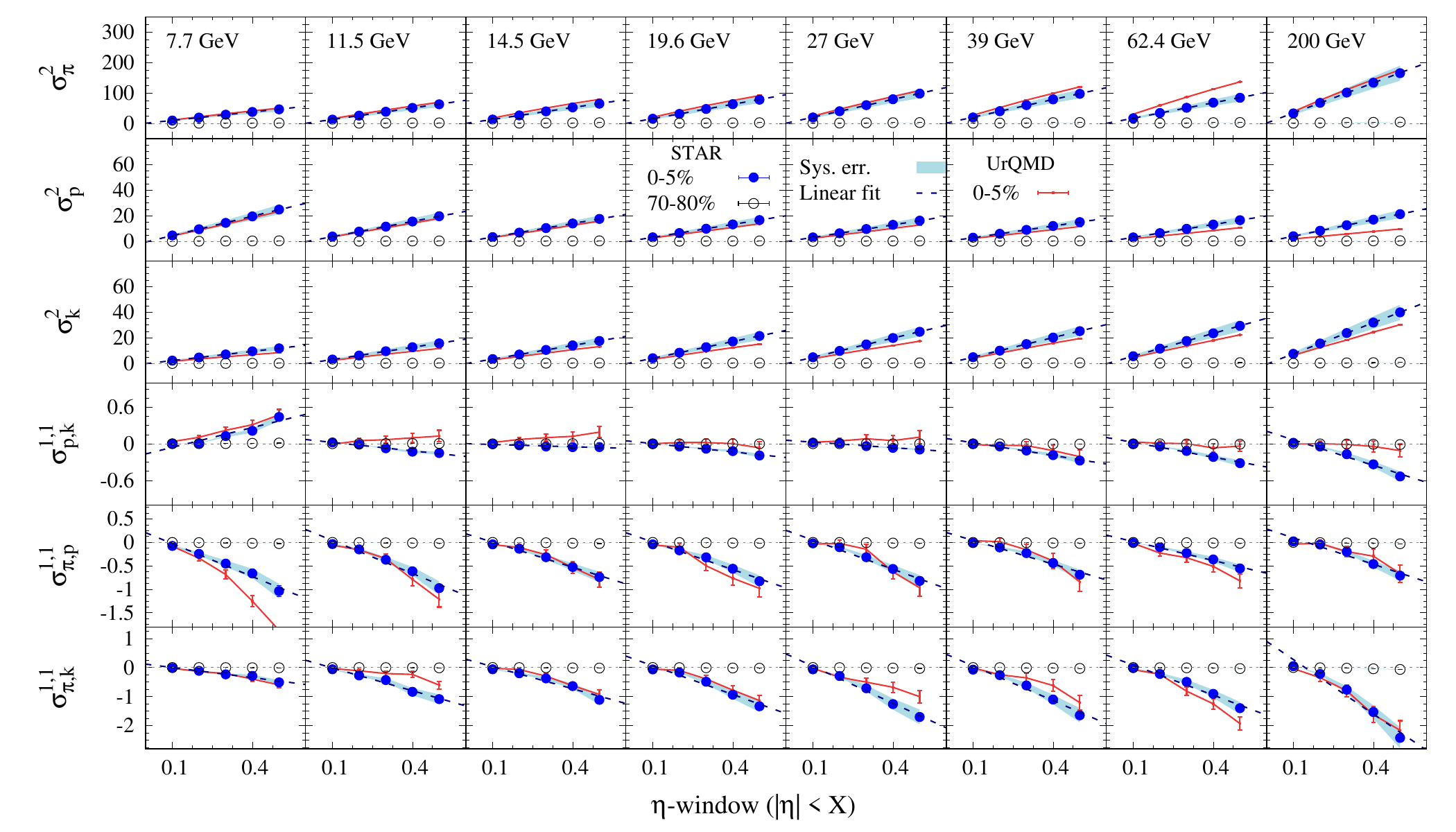}
  	\caption{The dependence of efficiency corrected second-order diagonal and off-diagonal cumulants on the width of the $\eta$-window. The filled and open circles represent 0-5$\%$ and 70-80$\%$ central collisions respectively. The shaded band represents the systematic uncertainty. The statistical uncertainties are within the marker size and solid lines are UrQMD calculations.}
  	\label{deltaeta} 
 \end{figure*}

In this erratum we present the following figures that are updated from the same in our original paper.\\

 Figure~\ref{deltaeta} shows the updated efficiency corrected diagonal and off-diagonal cumulants of net-pion, net-kaon and net-proton as a function of the $\eta$-window for the 0-5\% and 70-80\% centrality bins, and for eight collision energies. Here we replace the results of $\sigma^{2}_{Q}$, $\sigma^{1,1}_{Q,p}$ and $\sigma^{1,1}_{Q,k}$ with $\sigma^{2}_{\pi}$, $\sigma^{1,1}_{\pi,p}$ and $\sigma^{1,1}_{\pi,k}$, respectively. The results for \SigPK, $\sigma^{2}_{p}$, and $\sigma^{2}_{k}$ remain unchanged. The \SigpiK~and \SigpiP~ show a linearly decreasing trend with increasing pseudorapidity acceptance window ($\eta$-window).\\
 
\begin{figure*}[th]
 	\includegraphics[width=\textwidth]{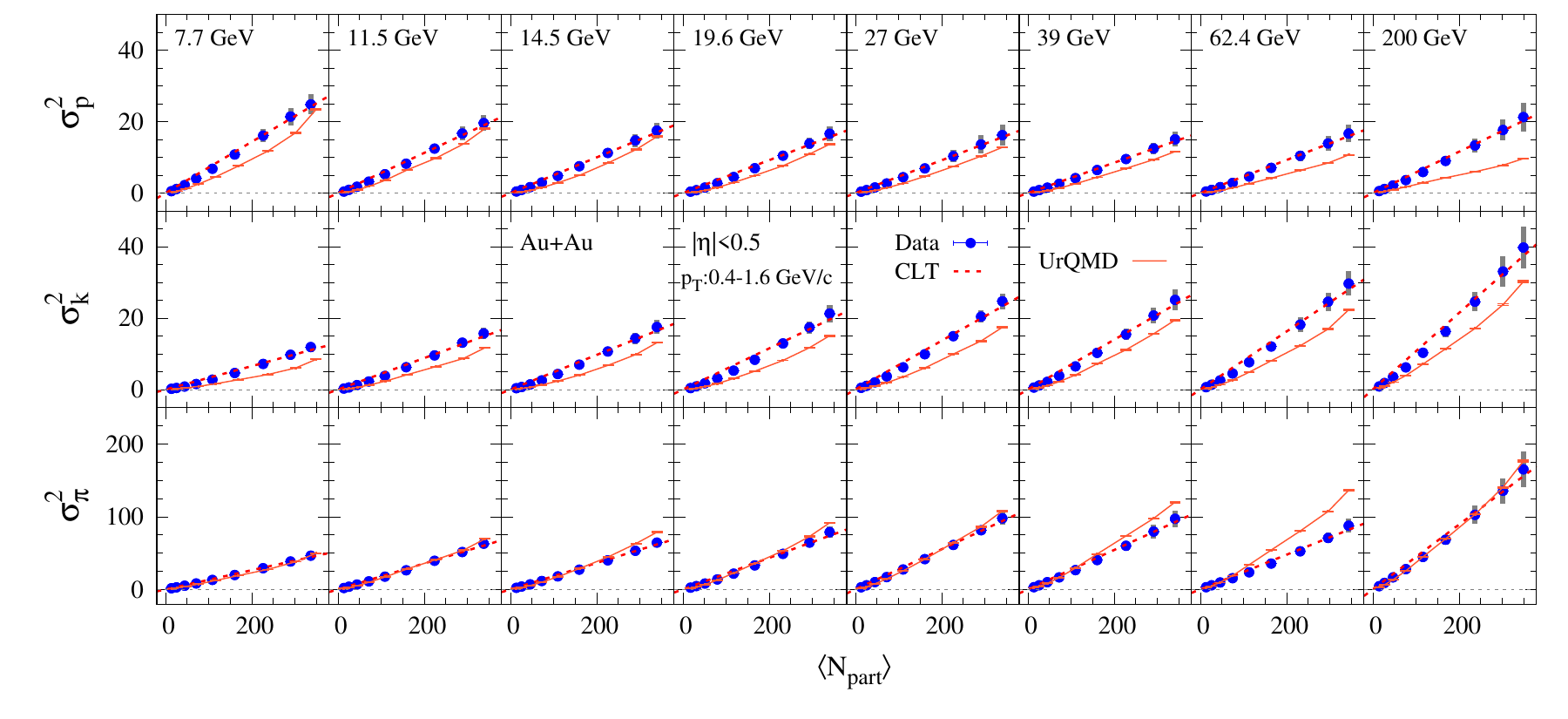}
 	\caption{Centrality dependence of efficiency corrected second-order diagonal cumulants of net-proton, net-kaon and net-pion (top to bottom) of the multiplicity distributions for \gold collisions at \sNN~= 7.7, 11.5, 14.5, 19.6, 27, 39, 62.4 and 200 GeV (left to right) within kinematic range of $|\eta|<0.5$ and $0.4<p_{\text{T}}<1.6$ GeV/\textit{c}. The boxes represent the systematic error. The statistical error bars are within the marker size. The dashed lines represent scaling predicted by the central limit theorem and the solid lines are UrQMD calculations.}
 	\label{variance}
 \end{figure*}
 
In Fig.~\ref{variance}, the $\sigma^{2}_{Q}$ is supplanted by $\sigma^{2}_{\pi}$ that shows a linear increasing trend as a function of collision centrality and agrees well with the UrQMD calculations. The results for $\sigma^{2}_{p}$ and $\sigma^{2}_{k}$ remain unchanged.\\

In Fig.~\ref{covariance}, the \SigQK~and \SigQP~ are replaced by \SigpiK~and \SigpiP~respectively. 
The values of \SigpiK~and \SigpiP~are negative at all collision energies, which indicates $\pi$-$p$ and $\pi$-$k$ are anti-correlated.\\ 

\begin{figure*}[th]
 	 	\includegraphics[width=\textwidth]{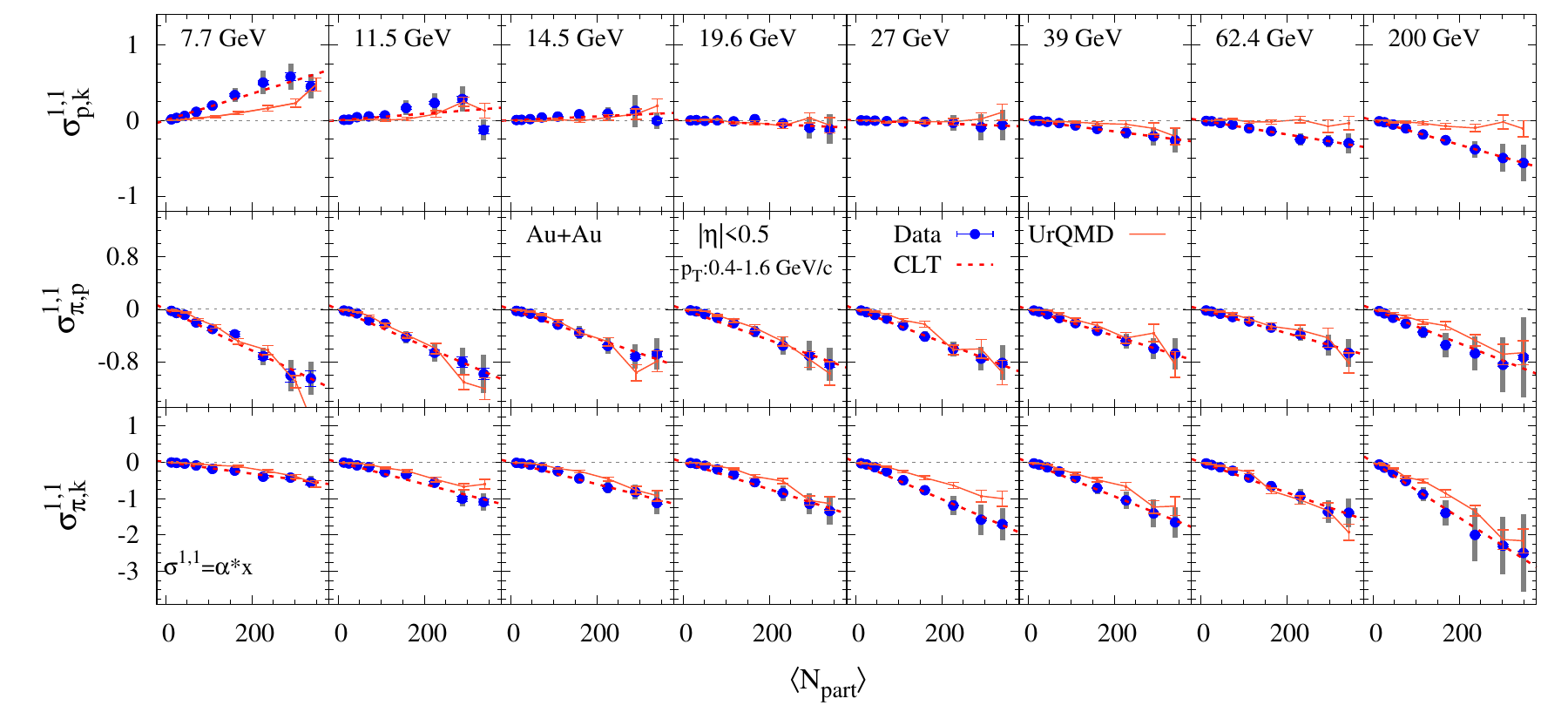}
 	\caption{Centrality dependence of second-order off-diagonal cumulants of net-proton, net-charge and net-kaon for \gold collisions at \sNN~= 7.7, 11.5, 14.5, 19.6, 27, 39, 62.4 and 200 GeV (left to right) within kinematic range $|\eta|<0.5$ and $0.4<p_{\text{T}}<1.6$ GeV/\textit{c}. Bars represent statistical errors and boxes show systematic errors. The dashed lines represent scaling predicted by the central limit theorem and the solid lines are UrQMD calculations.}
 	\label{covariance}
 \end{figure*}

  \begin{figure*}[t]
  	\includegraphics[width=\textwidth]{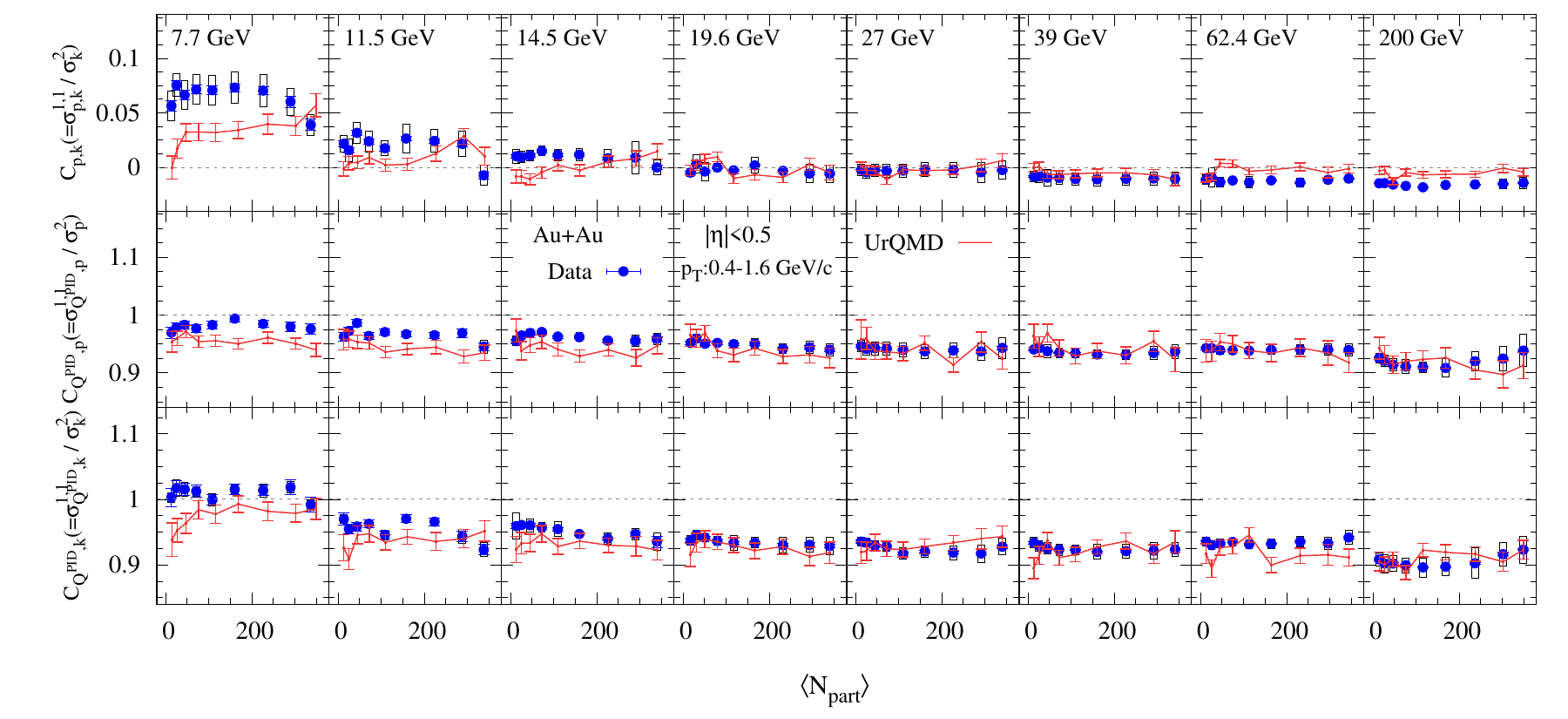}
  	\caption{Centrality dependence of second-order off-diagonal to diagonal cumulants ratios of net-proton, net-kaon and identified net-charge for Au+Au collisions at \sNN~= 7.7, 11.5, 14.5, 19.6, 27, 39, 62.4 and 200 GeV (left to right) within the kinematic range $|\eta|<0.5$ and $0.4<p_{\text{T}}<1.6$ GeV/\textit{c}. Bars represent statistical errors and boxes show systematic errors. The solid lines represent the UrQMD calculations.}
  	\label{ratio} 
     \end{figure*}

Figure~\ref{ratio} shows the centrality dependence of cumulant ratios. The quantities $C_{Q^{\rm PID},p}$ and $C_{Q^{\rm PID},k}$ are updated using Eq.~\ref{PIDQPratio} and Eq.~\ref{PIDQKratio}, respectively. The current data points agree with UrQMD.\\

Figure~\ref{energy} shows the collision energy dependence of $C_{p,k}$, $C_{Q^{\rm PID},p}$, and $C_{Q^{\rm PID},k}$  for 0-5\% and 70-80\% centralities.  The results are compared with UrQMD and HRG calculations.  The UrQMD calculations are redone using Eq.~\ref{PIDQPratio} and Eq.~\ref{PIDQKratio}.  The quantities $C_{Q^{\rm PID},p}$ and $C_{Q^{\rm PID},k}$ decrease with collision energy and are below the Poisson baseline. The quantity $C_{Q^{\rm PID},k}$ agrees well with both the UrQMD and HRG calculations. The quantity $C_{Q^{\rm PID},p}$ agrees with the UrQMD calculations but deviates from the HRG results.\\

  \begin{figure}[t]
  	\includegraphics[width=0.48\textwidth]{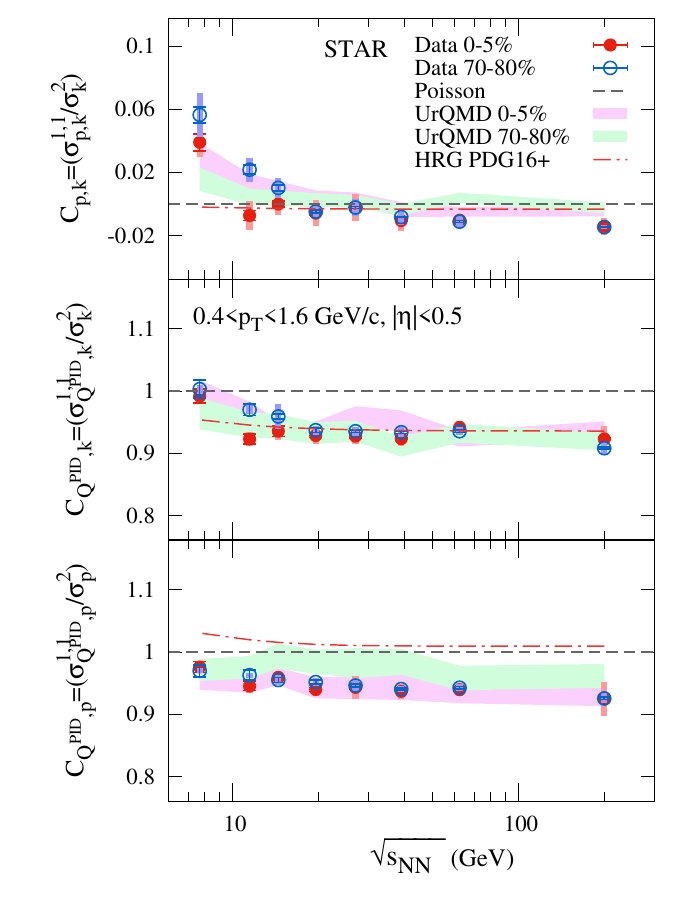}
\caption{Beam energy dependence of the cumulant ratios ($C_{p,k}$, $C_{Q^{\rm PID},k}$, and $C_{Q^{\rm PID},p}$, top to bottom) of net-proton, net-kaon, and identified net-charge for Au+Au collisions at $\sqrt{s_{\rm NN}}$ = 7.7, 11.5, 14.5, 19.6, 27, 39, 62.4, and 200 GeV.  The bands denote the UrQMD calculations for 0-5\% and 70-80\% centralities. The HRG calculations are represented by red dotted lines. The Poisson baseline is given by the black dashed lines. Bars show statistical errors and boxes show systematic errors.}
\label{energy} 
   \end{figure}


In summary, we address the issue of self-correlation in the previously considered efficiency correction for: 1) net-charge and net-proton and 2) net-charge and net-kaon second-order off-diagonal cumulants. For these quantities, we replace unidentified charged hadrons, as used in our original paper~\cite{Adam:2019xmk}, with the sum of pions, kaons, and protons. Unlike our previous observations reported in Ref.~\cite{Adam:2019xmk}, we see the following differences: 1) the cumulant ratios do not show strong dependence on centrality or collision energy, 2) for the cumulant ratio of identified net-charge and net-kaon ($C_{Q^{\rm PID},k}$) we do not see any strong deviation from UrQMD or HRG calculations, 3) for the identified net-charge and net-proton case ($C_{Q^{\rm PID},p}$), we observe that the results are slightly below the HRG calculations but are consistent with the UrQMD calculations.

  \section*{ACKNOWLEDGMENTS}
  
We thank V. Vovchenko and V. Koch for pointing out the issues of efficiency correction with off-diagonal cumulants and important discussions. We thank the RHIC Operations Group and RCF at BNL, the NERSC Center at LBNL, and the Open Science Grid consortium for providing resources and support.  This work was supported in part by the Office of Nuclear Physics within the U.S. DOE Office of Science, the U.S. National Science Foundation, the Ministry of Education and Science of the Russian Federation, National Natural Science Foundation of China, Chinese Academy of Science, the Ministry of Science and Technology of China and the Chinese Ministry of Education, the Higher Education Sprout Project by Ministry of Education at NCKU, the National Research Foundation of Korea, Czech Science Foundation and Ministry of Education, Youth and Sports of the Czech Republic, Hungarian National Research, Development and Innovation Office, New National Excellency Programme of the Hungarian Ministry of Human Capacities, Department of Atomic Energy and Department of Science and Technology of the Government of India, the National Science Centre of Poland, the Ministry  of Science, Education and Sports of the Republic of Croatia, RosAtom of Russia and German Bundesministerium fur Bildung, Wissenschaft, Forschung and Technologie (BMBF), Helmholtz Association, Ministry of Education, Culture, Sports, Science, and Technology (MEXT) and Japan Society for the Promotion of Science (JSPS).


\clearpage

\section*{Collision energy dependence of second-order off-diagonal and diagonal cumulants of net-charge, net-proton and net-kaon multiplicity distributions in Au+Au collisions}

\section*{abstract}
	
	We report the first measurements of a complete second-order cumulant matrix of net-charge, net-proton and net-kaon multiplicity distributions for the first phase of the beam energy scan program at RHIC. This includes the centrality and, for the first time, the pseudorapidity window dependence of both diagonal and off-diagonal cumulants in Au+Au collisions at \sNN~= 7.7-200 GeV. Within the available acceptance of $|\eta|<0.5$, the cumulants grow linearly with the pseudorapidity window. 
	Relative to the corresponding measurements in peripheral collisions, the ratio of off-diagonal over diagonal cumulants in central collisions indicates an excess correlation between net-charge and net-kaon, as well as between net-charge and net-proton.
	The strength of such excess correlation increases with the collision energy. 
	The correlation between net-proton and net-kaon multiplicity distributions is observed to be negative at \sNN~= 200 GeV and change to positive at the lowest collision energy.
	Model calculations based on non-thermal (UrQMD) and thermal (HRG) production of hadrons cannot explain the data.
	These measurements will help map the QCD phase diagram, constrain hadron resonance gas model calculations and provide new insights on the energy dependence of baryon-strangeness correlations. 
	

\section{Introduction}

Ever since the first discussion of possible signatures of the quark-gluon plasma (QGP)~\cite{Collins:1974ky,Chin:1978gj,Kapusta:1979fh,Anishetty:1980zp} at the Relativistic Heavy-Ion Collider (RHIC)~\cite{Arsene:2004fa, Back:2004je, Adams:2005dq, Adcox:2004mh}, physicists have been exploring the landscape of the Quantum Chromodynamics (QCD) phase diagram and trying to locate the conjectured critical endpoint (CP)~\cite{Berges:1998rc,Halasz:1998qr}. About a decade ago, the Beam Energy Scan (BES) program was proposed at the RHIC to achieve such a goal by colliding heavy ions over a wide range of beam energies~\cite{Aggarwal:2010cw}. 
One of the primary aims of such a program was to identify the signature of criticality in the measurements of 
%
%
event-by-event fluctuations of the net-multiplicity ($\delta N$) of different particle species that carry different conserved charges ($\alpha$) such as net-electric charge ($Q$), net-baryon number ($B$), and net-strangeness ($S$). 
It is suggested that the $n$-th order cumulants of the net-multiplicity distributions ($\kappa^{n}_{\alpha}[\delta N]$) are related to the $n$-th order thermodynamic susceptibilities ($\chi^{n}_{\alpha}$) of the corresponding conserved charges in QCD that diverge near the CP~\cite{Stephanov:2008qz, Cheng:2008zh, Asakawa:2009aj,Stephanov:2011pb,Friman:2011pf}. Therefore, measurements of $\kappa^{n}_{\alpha}[\delta N]$ can be used to signal the presence of the CP~\cite{Stephanov:1999zu,Stephanov:2008qz}. The STAR and PHENIX experiments, over past few years, have measured such higher-order cumulants of the net-charge ($Q$)~\cite{Adamczyk:2014fia,Adare:2015aqk}, net-proton ($p$, a proxy for the net-baryon)~\cite{Aggarwal:2010wy, Adamczyk:2013dal}, and net-kaon ($k$, a proxy for the net-strangeness)~\cite{Adamczyk:2017wsl} multiplicity distributions, although no distinctive signatures of the CP have been inferred from such measurements. 
%
%
In addition, these measurements 
have also been used to extract the freeze-out temperature ($T$) and baryon chemical potential ($\mu_{\text{B}}$), at a given collision energy, by comparing the data with hadron resonance gas model (HRG) and lattice QCD calculations~\cite{Adamczyk:2014fia,Adare:2015aqk,Bazavov:2012vg,Borsanyi:2013hza,Alba:2014eba, Noronha-Hostler:2016rpd}.

%
%

So far, RHIC measurements have focused on diagonal cumulants ($\kappa^{n}_{\alpha}$) which quantify the self-correlation of a specific kind of conserved charge ($\alpha$).
Similar to the diagonal cumulants, one can readily construct and measure off-diagonal cumulants ($\kappa^{m,n}_{\alpha,\beta}$) of the net-charge, net-proton, and net-kaon multiplicity distributions in heavy-ion experiments. As we alluded to previously, these off-diagonal cumulants are related to the off-diagonal thermodynamic susceptibilities ($\chi^{m,n}_{\alpha,\beta}$) that carry the correlation between different conserved charges ($\alpha,\beta$) of QCD~\cite{Koch:2005vg,Gavai:2005yk,Majumder:2006nq,Bluhm:2008sc, Ding:2015fca}. 
%
%
%
%
%
%
%
%
%
The importance of studying off-diagonal cumulants was first highlighted in the context of baryon-strangeness correlations ~\cite{Koch:2005vg}, which can be studied by measuring the energy dependence of the ratios of off-diagonal over diagonal cumulants $\kappa^{1,1}_{B,S}/\kappa^{2}_{S}$. Such ratios can be quantified by the susceptibility ratio $C_{B,S} = -3\chi^{1,1}_{B,S}/\chi^2_{S}$ and are expected to show a rapid change with the onset of deconfinement~\cite{Koch:2005vg,Majumder:2006nq,Bazavov:2013dta,Bazavov:2014xya}.


%
%
%
%

Another impetus for studying off-diagonal cumulants comes from the comparisons of lattice QCD and ideal HRG model calculations~\cite{Bazavov:2012jq,Vovchenko:2016rkn}. 
One expects ideal HRG to be a good approximation of QCD matter below the crossover transition temperature (e.g. $T_c$ = 154 ($\pm$9) MeV, at $\mu_B$ = 0~\cite{Bazavov:2011nk}). 
However, the baryon-charge susceptibility $\chi^{1,1}_{B,Q}$ shows a significant difference between ideal HRG and lattice calculations~\cite{Bazavov:2012jq,Vovchenko:2016rkn}. A similar difference between HRG and lattice can also be seen in higher-order baryon susceptibilities ($\chi^{4}_B$). It turns out that the off-diagonal cumulants, even at the level of second-order, show significant sensitivity to the difference between the calculations from the ideal HRG and lattice~\cite{Karsch:2017zzw}. 
%
%
Calculations presented in ~\cite{Vovchenko:2016rkn} demonstrated that by including additional interactions among hadrons it may be possible to explain the difference between lattice and HRG calculations for $\chi^{1,1}_{B,Q}$.  
%
%
%
%
Therefore, measurements of off-diagonal moments will help constrain different hadron gas models that include various assumptions on the underlying baryon-meson interactions, species dependent freeze-out temperatures, and the number of resonance states~\cite{Vovchenko:2016rkn,Bellwied:2018tkc,Bellwied:2013cta,Chatterjee:2013yga,Karsch:2010ck}. The measurements of off-diagonal cumulants will enable independent extraction of freeze-out parameters, as obtained previously using diagonal cumulants.


%


%
%

It is important to take into account the sensitivity of the off-diagonal and diagonal cumulants to the experimental inefficiency of detecting neutral and heavy particles that also carry conserved charges. %
In most heavy-ion experiments, the measurements of the total number of produced baryons are challenged by the lack of detection capability of neutral baryons (e.g. neutrons). The same is also true for the measurements of strange particles. 
It is difficult to perform high-purity event-by-event measurements of neutral strange baryons such as $\Lambda$, strange mesons such as $K_S^{0}$ or other heavy conserved charge-carrying particles such as $\Omega, \Sigma, \Xi$, etc.
This is because they require reconstruction using invariant mass spectra that reduces both the efficiency and purity of their detection~\cite{Adamczyk:2017nxg}. 
%
One, therefore, uses the number of net-protons ($p$) and net-kaons ($k$) as proxies for the measurements of $\kappa^{n}_{B}$ and $\kappa^{n}_{S}$. Only the measurement of $\kappa^{n}_{Q}$ does not require any proxy. 
On the other hand, measurements of off-diagonal cumulants such as $\kappa^{m,n}_{Q,B}$ or $\kappa^{m,n}_{Q,S}$ are less affected by the experimental inability to measure neutral baryons or neutral strange particles, as they do not contribute to such correlations. They can be approximated as $\kappa^{m,n}_{Q,B} \approx \kappa^{m,n}_{Q,p}$ and $\kappa^{m,n}_{Q,S}\approx \kappa^{m,n}_{Q,k}$~\cite{Chatterjee:2016mve}. 
Without measuring strange-baryons, one cannot simply approximate $\kappa^{m,n}_{B,S}$ by $\kappa^{m,n}_{p,k}$.
However, one expects a reasonable connection between the two quantities~\cite{Chatterjee:2016mve,Yang:2016xga}. Measurement of $\kappa^{m,n}_{p,k}$ therefore provides access to essential albeit qualitative features of a rapid change of baryon-strangeness correlations near deconfinement transition as predicted in~\cite{Koch:2005vg}.
%

%
%
%
%
%
%
%

%
We present the measurements of the second-order diagonal and off-diagonal cumulants of net-charge, net-proton, and net-kaon distributions within the common acceptance in Au+Au collisions at \sNN~= 7.7, 11.5, 14.5, 19.6, 27, 39, 62.4 and 200 GeV from the STAR experiment. We show a comparison of our results with hadronic models, including HRG and UrQMD~\cite{Bass:1998ca, Bleicher:1999xi}. 

The paper is organized as follows. In the next section (section \ref{sce:oberservables}) we define the observables and notations used in this analysis. In section \ref{sce:experiment}, we discuss experimental details and analysis techniques including particle identification, centrality selection, centrality bin-width correction, efficiency correction, and uncertainty estimation. We discuss the results in section~\ref{sce:results} and summarize in section~\ref{sec:summary}.

\section{OBSERVABLES}\label{sce:oberservables}
Different second-order thermodynamic number susceptibilities of the conserved charges at thermal and chemical equilibrium are related to the corresponding second-order diagonal and off-diagonal cumulants of net-multiplicity distributions~\cite{Cheng:2008zh} as,
\bea
\chi^{2}_{\alpha} = \frac{1}{VT^{3}} \kappa^{2}_{\alpha}, \,\,\,\,\,\,
\chi^{1,1}_{\alpha,\beta} = \frac{1}{VT^{3}} \kappa^{1,1}_{\alpha,\beta},
\eea
where $V$ and $T$ are the system volume and temperature. The second-order cumulants, also referred to as the variance ($\sigma^{2}_{\alpha}$) and covariance ($\sigma^{1,1}_{\alpha,\beta}$), respectively, can be expressed as 
\bea
\kappa^{2}_{\alpha}= \sigma^{2}_{\alpha} =  \la (\delta N_{\alpha} - \la \delta N_{\alpha} \ra)^{2} \ra 
\eea
and
\bea
\kappa^{1,1}_{\alpha,\beta}= \sigma^{1,1}_{\alpha,\beta} = \la (\delta N_{\alpha} - \la \delta N_{\alpha} \ra) (\delta N_{\beta} - \la \delta N_{\beta} \ra) \ra.
\eea
Here, $\la\cdots\ra$ represents an average over the events with $\delta N_{\alpha} = N_{\alpha^{+}}-N_{\alpha^{-}}$ and $\alpha$, $\beta$ can be $p$, $Q$ and $k$ for the current measurements. 
%
%
It is more convenient to write all possible combinations of cumulants in a matrix form as  
\be
\sigma =
\left(
\begin{array}{ccc}
	\sigma^{2}_{Q} &  \sigma^{1,1}_{Q,p}  &   \sigma^{1,1}_{Q,k}\\\\
	\sigma^{1,1}_{p,Q} &  \sigma^{2}_{p}  &   \sigma^{1,1}_{p,k}\\	\\
	\sigma^{1,1}_{k,Q} &  \sigma^{1,1}_{k,p}  &   \sigma^{2}_{k}\\
\end{array}
\right).
\ee

Since $\sigma^{1,1}_{\alpha,\beta} = \sigma^{1,1}_{\beta,\alpha}$, we present measurements of the six independent components of this cumulant matrix at the different beam energies, centralities and windows of pseudorapidity. 


\section{Dataset and experimental details}\label{sce:experiment}
We make use of the data from Au+Au collisions at RHIC collected by the STAR detector~\cite{Ackermann2003624} over the years 2010 to 2014. We analyze minimum-bias (MB) events for eight different energies, \sNN~= 7.7, 11.5, 14.5, 19.6, 27, 39, 62.4 and 200 GeV, acquired by requiring the coincidence of signals from the Zero Degree Calorimeters (ZDCs)~\cite{Adler:2000bd} and the Vertex Position Detectors (VPDs)~\cite{Llope:2003ti}.  
%
%
%
%
STAR has uniform acceptance at mid-rapidity of $|\eta|<1$, a full $2\pi$ azimuthal coverage, and excellent particle identification. The Time Projection Chamber (TPC) ~\cite{Anderson:2003ur} sits inside a 0.5 T magnet and records the charged particle tracks, measures their momenta, and identifies them based on their energy loss ($dE/dx$). We use the TPC to reconstruct the position of the primary vertices of collisions along the beam direction ($V_{z}$) and along radial direction transverse to the beam axis ($V_r$).
%
%
For the current analysis we restrict the positions of primary vertices to be $|V_{z}|<30$ cm and $V_{r}<2$ cm. 
RHIC delivers collisions at higher luminosity for higher energies \sNN~= 39, 62.4 and 200 GeV that increases the probability of pile-up events. In order to suppress such pile-up events we apply an additional cut on the absolute difference between the \textit{z}-vertex positions determined by two different detectors (TPC and VPD),
{\it i.e.} $\left | V_{\textit{z}} \rm(VPD) - \it V_{\textit{z}}\rm(TPC)\right| < 3$ cm. 
%
%
%
In addition, pile-up events have been removed by taking correlation between the number of TPC tracks and number of TOF matched tracks. 

For the calculation of cumulants, we use charged tracks reconstructed by the TPC within $|\eta|< 0.5$, and with transverse momentum $0.4< p_{\rm{T}}<1.6$ GeV/\textit{c}. To reduce the contamination from the secondary charged particles, we only select tracks with a distance of closest approach (DCA) from the primary vertex less than 1 cm. 
We also require at least twenty ionization points (nFitPoints) in the TPC for selecting a good track. 
%

%


\subsection{Particle identification}
\begin{figure}[t]
	\centering 	 	
	\includegraphics[width=0.5\textwidth]{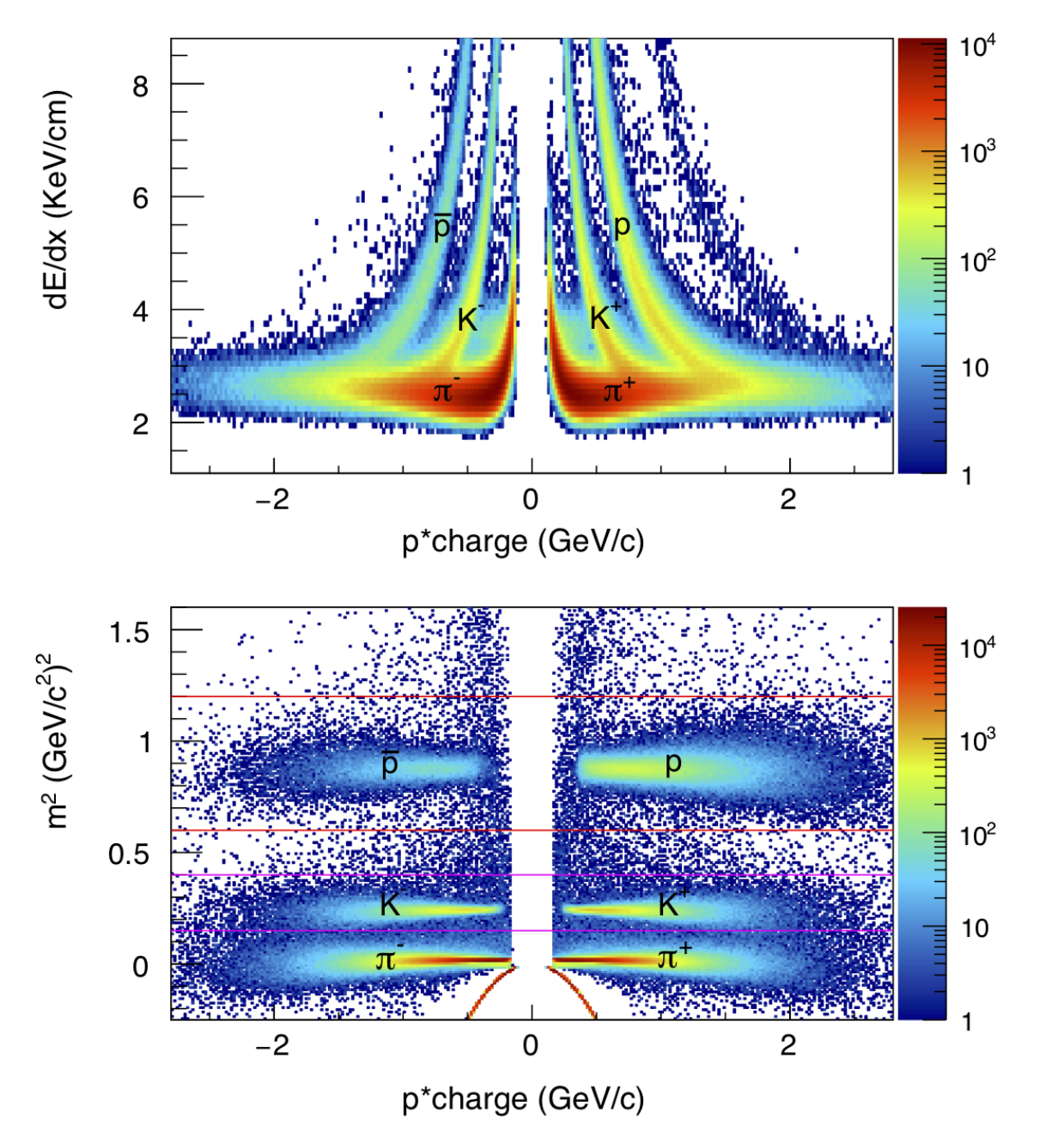}
	\caption{Top: $dE/dx$ from TPC plotted against charge $\times$ momentum of individual particles for Au+Au collisions at \sNN~= 27 GeV. Bottom: $m^{2}$ from the TOF detector plotted as a function of charge $\times$ momentum for Au+Au collisions at \sNN~= 27 GeV. The red and pink lines represent the proton and kaon selection cuts respectively. Similar distributions are obtained for all other collision energies.} 
	\label{fig_pid} 
\end{figure}

%
%
We use a combination of the TPC and Time-of-Flight (TOF)~\cite{Llope:2003ti} detectors for the measurements of (anti-) protons ($p(\bar{p})$) and (anti-) kaons ($K^\pm$) within the same acceptance. 
%
Figure~\ref{fig_pid} (top) shows the distribution of the energy loss of charged tracks passing through the TPC, plotted against charge times momentum.
%
%
To achieve a good purity in the sample of identified particle species ``$X$", we determine a quantity 
$n\sigma_{X}$ defined as, 
\bea
n\sigma_{X} = \frac{\rm {ln}[ (dE/dx)_{\rm Measured}/(dE/dx)_{\rm Bichsel}]}{\sigma_{X}}.
\eea
Here $(dE/dx)_{\rm Measured}$ is the ionization energy loss measured by the TPC, and $(dE/dx)_{\rm Bichsel}$ is the corresponding theoretical value from Bichsel curves estimated for each identified particle using an extension of the Bethe-Bloch formula~\cite{Abelev:2008ab}. The quantity $\sigma_{X}$ is the $dE/dx$ resolution of TPC. %
%
%
It is obvious from Fig.~\ref{fig_pid} that the identification using TPC is limited to low momenta where distinct $dE/dx$ bands are observed for different particle species. 
We, therefore, use TOF to improve particle identification over a wider range of momenta by measuring the flight time ($t$) of a particle from the primary vertex of a collision. 
By combining such information with the path length ($L$) traversed by the particle, measured by TPC, one can directly calculate the velocity ($v$) and mass ($m$) using the expressions:
\bea
\beta = \frac{v}{c} &=& \frac{L}{ct}, \\
m^{2} &=& p^{2}\left( \left( \frac{1}{\beta} \right)^{2} - 1 \right).
\eea

%
%
Fig.~\ref{fig_pid} (bottom) shows the distribution of the $m^2$ against charge times momentum. This is used to identify different particle species. This additional information of $m^2$ helps us to identify $p(\bar{p})$ and $K^\pm$ in the region of higher momentum where their $dE/dx$ distributions merge as shown in Fig.~\ref{fig_pid} (top). 
%
%
%
%
More specifically, for particles with $0.4<p_{\text{T}}<0.8$ GeV/\textit{c} we use the TPC to identify the $p(\bar{p})$ using a cut of $| n\sigma_{p} |  < 2$. To identify $p(\bar{p})$ in the range $0.8<p_{\text{T}}<1.6$ GeV/\textit{c}, we apply an additional cut of $0.6< m^{2} < 1.2$ GeV$^{2}$/\textit{c}$^{4}$ using TOF. In case of $K^\pm$, we use the following criteria: $0.15 < m^{2} < 0.4$ GeV$^{2}$/\textit{c}$^{4}$, $| n\sigma_{K} | < 2$ and $| n\sigma_{p} | > 2$ for the entire range of transverse momentum, {\it i.e.}, $0.4 <p_{\text{T}}< 1.6$ GeV/\textit{c}. The purities of $ K^{\pm}$ and p({$\bar{p}$) are found to be 98\% and 99\%, respectively.
	%
	
	
	
	%
	%
	
	\subsection{Centrality determination and bin-width correction}

	In order to determine collision centrality we use the distribution of the measured charged-particle multiplicity ($N_{\rm trk}$) within $0.5 < |\eta| < 1$. Thus, we exclude the particles used to calculate the cumulants from the particles used to determine the centrality to reduce autocorrelation effects~\cite{Adamczyk:2013dal, Adamczyk:2014fia}. We perform our analysis for nine centrality intervals (0-5\%, 5-10\%, 10-20\%, ..., 70-80\%), and use a Monte Carlo Glauber model~\cite{Abelev:2008ab,Miller:2007ri} to estimate the average number of participating nucleons $N_{\rm part}$ for each of these intervals. For details, we refer the reader to~\cite{Abelev:2008ab}. 
	
	
	The conventional approach to centrality analysis leads to an artifact in the event-by-event analysis of cumulants known as the centrality bin width (CBW) effect \cite{Sahoo:2012wn, Luo:2013bmi}. 
	This happens because a given centrality class (e.g. $0$-$5\%$) is determined using the charged-particle multiplicity (uncorrected) distribution. 
	A particular window of $N_{\rm trk}$ corresponds to a large variation of impact parameter and collision geometries. Such variations lead to volume fluctuations, complicating the picture of ensemble averaging over identical configurations. Also, cumulants of  different orders can have different sensitivity to such fluctuations~\cite{Braun-Munzinger:2016yjz}. In principle, CBW cannot be removed completely due the lack of knowledge of the collision geometry in heavy-ion collisions. These effects can be  minimized by choosing narrowest possible windows of $N_{\rm trk}$. In order to both minimize CBW and present the final results in terms of conventional centrality intervals ($0$-$5\%$, 5-10\%, 10-20\%, ..., 70-80\%), we perform the following procedure. 
	%
	%
	%
	{
		%
		%
		%
		%
		
		We first estimate different cumulants in bins of unit multiplicity and then weight the cumulants by the number of events in each bin over a desired centrality class. This can be expressed as, 
		\bea
		{\cal O} =  \frac{ \sum_{i} n_{i} {\cal O}_{i}}{\sum_{i} n_{i}} = \sum_{i} \omega_{i} {\cal O}_{i}.
		\eea
		Here ${\cal O}_{i}$ is the observable measured in the $i^{th}$ multiplicity bin, $n_{i}$ and $\omega_{i}$ $(= \frac{ n_{i}}{\sum_{i} n_{i}} )$ are the number of events and the weight factor for  $i^{th}$ multiplicity bin, respectively. This approach was implemented in previous publications from STAR and PHENIX~\cite{Adamczyk:2014fia,Adare:2015aqk,Adamczyk:2013dal}. A number of independent studies indicate that the CBW effect is 
		negligible for lower-order ($\leq 2$) cumulants~\cite{Sahoo:2012wn, Luo:2013bmi}. Note that statistical uncertainties of the cumulants also require the same CBW correction. All the results presented in this paper include CBW correction.}
	
	\subsection{Efficiency correction}
	Cumulant measurements are complicated by the finite efficiency of detection.  
	%
	We perform the efficiency correction in two steps: first, we determine the numerical values of the efficiency using detector simulation and then we use the algebra based on binomial detector response~\cite{Bzdak:2012ab} to correct the measurements of individual cumulants. A major challenge in this context arises from the dependence of efficiency on particle species and transverse momentum which leads to a cumbersome algebra of efficiency correction~\cite{Bzdak:2013pha}.
	%
	%

	For the first step, we estimate the tracking efficiency using simulations based on the \textsc{geant}~\cite{Fine:2000gn} implementation of the TPC. The efficiency values of proton and anti-proton, for all beam energies, vary between 60-80\% and 80-83\% at the most central (0-5\%) and peripheral (70-80\%) centralities, respectively, at low-\pT~($0.4<p_{\text{T}}<0.8$ GeV/\textit{c}). 
	As mentioned above, we use a combination of TPC and TOF for the identification of high-\pT~particles. We estimate the combined TPC+TOF efficiency for high-\pT~particles by multiplying the TPC tracking efficiency and TOF matching efficiency. 
	The TOF matching efficiency is estimated by comparing the number of tracks that are detected in TPC and the ones that also have corresponding hits in TOF. The combined TPC+TOF efficiency is approximately 30\% lower than the TPC tracking efficiency because not every track detected in TPC can be matched to a corresponding hit in TOF.  
	For $p(\bar{p})$, the TPC+TOF efficiency varies between 40-60\% at all centralities and beam energies within  $0.8<p_{\text{T}}<1.6$ GeV/\textit{c}. Similarly, for $K^\pm$, the TPC+TOF efficiency varies about 38-42\% in the range $0.4 < p_{\text{T}} < 1.6 $ GeV/\textit{c}. 
	In case of inclusive charged particles, measured within $0.4 < p_{\text{T}} < 1.6 $ GeV/\textit{c}, TOF-matching is not required. For $Q^{\pm}$, we find a variation of the efficiency between 60-80\% and 75-80\% at 0-5\% and 70-80\% centralities, respectively, for all eight energies. 
	
	For the second step, we apply the efficiency values in the algebraic expressions that relate the true cumulants to the measured ones. Such expressions are obtained by assuming an ansatz of binomial detector response~\cite{Bzdak:2012ab,Bzdak:2013pha,Luo:2014rea}.  
	%
	%
	The same approach of efficiency correction has been performed in the previous measurements of the diagonal cumulants~\cite{Adamczyk:2013dal, Adamczyk:2014fia, Adamczyk:2017wsl}. 
	It has been argued that deviations from binomial detector response will further complicate the efficiency corrections~\cite{Bzdak:2016qdc}. The effects of non-binomial detector response are currently being explored in the STAR collaboration~\cite{Adamczyk:2017wsl}. 
	Nevertheless, in a recent publication it has been explicitly demonstrated, using \textsc{hijing}+\textsc{geant} simulations with STAR geometry, that binomial detection response for efficiency correction can reproduce the cumulants of the initial input multiplicity distributions~\cite{Adamczyk:2017wsl,Fine:2000gn}. Particularly, for second-order cumulants, the binomial detector response is shown to be a reasonable approximation. 

	In this analysis, we apply binomial efficiency corrections for all six cumulants in two \pT~bins, nine centrality bins, and separately for particles and anti-particles. It must be noted that the statistical uncertainties of these cumulants have to be also corrected for detection efficiency~\cite{smomentPT}. A detailed discussion of both the statistical and systematic uncertainties can be found in the following section.

	\subsection{Uncertainty estimation}
	\label{error} 
	We estimate statistical uncertainties of the diagonal and off-diagonal cumulants using the analytical error propagation method~\cite{kendall1963advanced, Luo:2011tp}. 
	Statistical uncertainty of the cumulant of net-distributions depends on the variance of the distribution and the number of events ($n$). For a cumulant of any order, the statistical uncertainty is expressed in terms of higher-order cumulants. 
	Therefore, along with the cumulants, we also perform efficiency corrections to the estimated statistical uncertainties~\cite{Luo:2014rea}. 
	
	We estimate systematic uncertainties in our measurements by varying track selection criteria (DCA, nFitPoints values) and the conditions for particle identification ($|n\sigma_{K}|, |n\sigma_{p}\big|$ values). When we vary these cuts, we make sure the measured particle yields lie within $5\%$ of what is obtained for the default cuts. We take into account the correlations of the statistical uncertainties while studying the systematic effects.
	The feed down from weak decays decreases the purity of the proton and kaon samples, however in our case they are largely suppressed by applying DCA cuts. 
	We vary the DCA cut within a range of 0.8-1.2 cm and find that the magnitude of the cumulants  at \sNN~= 200 (7.7) GeV changes by about $10~(6)\%$. However, the variation of such cuts on the ratio of off-diagonal over diagonal cumulants is about $1\%$. 
	
	The variation of nFitPoints over a range of 16-24 leads to about 2\% variations in the cumulants. The particle identification condition and detection efficiency contribute among the dominant sources of ($5$-$7\%$) systematic uncertainty. 
	%
	%
	We also estimate systematic variations in the cumulant values by varying the tracking efficiency by $5\%$; such variations account for the uncertainty in the \textsc{geant} simulation. In this analysis, we find statistical uncertainties to be smaller (less than $5\%$) than the corresponding systematic uncertainties. We also find the systematic uncertainties have a weak dependence on beam energy.
	Overall systematic uncertainties lie within $8$-$15\%$ for all the results.  

	\begin{figure*}[th]
		\centering
		\includegraphics[width=\textwidth]{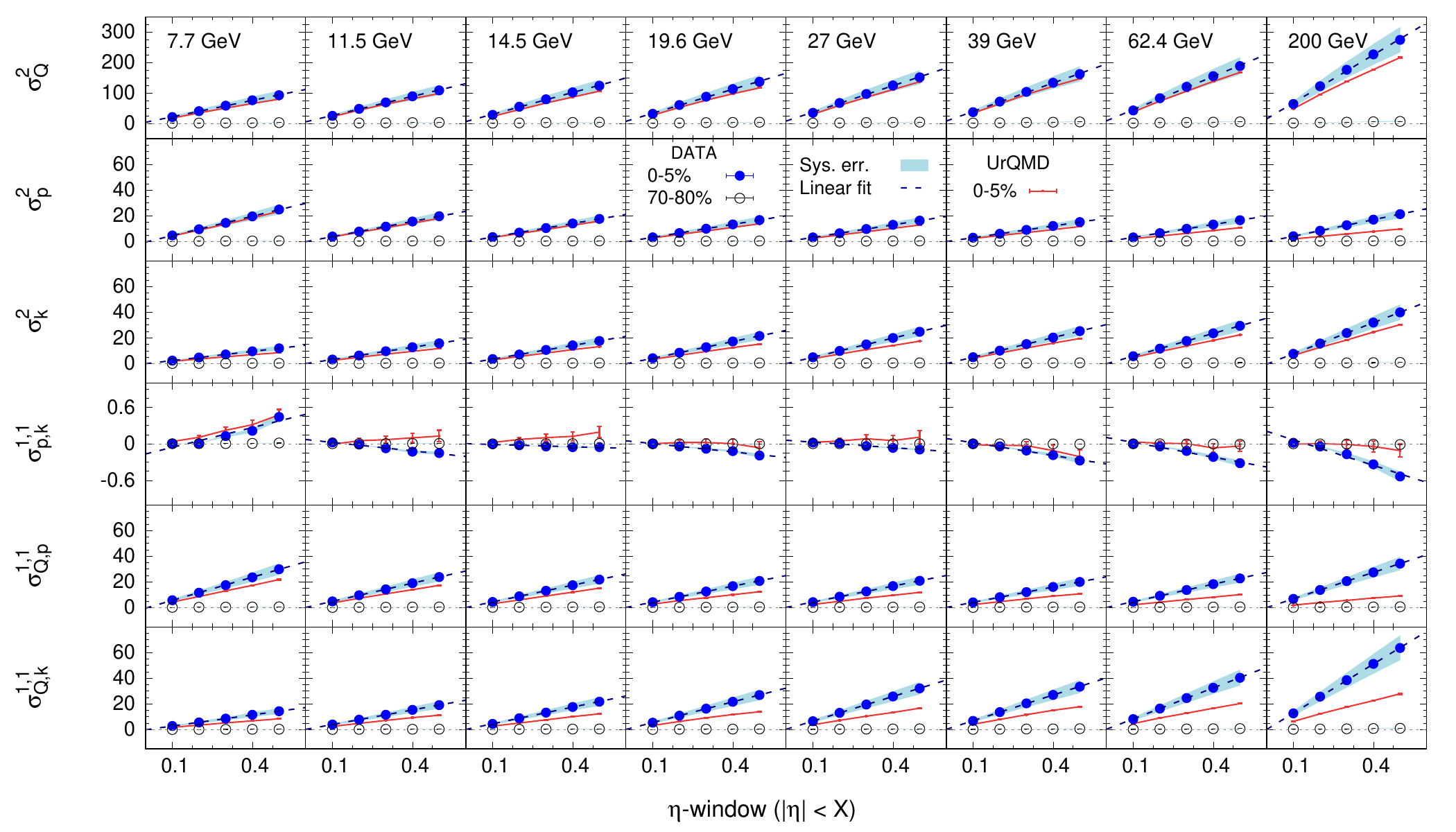}
		\caption{The dependence of efficiency-corrected second-order diagonal and off-diagonal cumulants on the width of the $\eta$-window. The filled and open circles represent 0-5$\%$ and 70-80$\%$ central collisions respectively. The shaded band represents the systematic uncertainty. The statistical uncertainties are within the marker size and solid lines are UrQMD calculations.}
		\label{deltaeta} 
	\end{figure*}

	\begin{figure*}[th]
		\vspace{2cm} 
		\includegraphics[width=\textwidth]{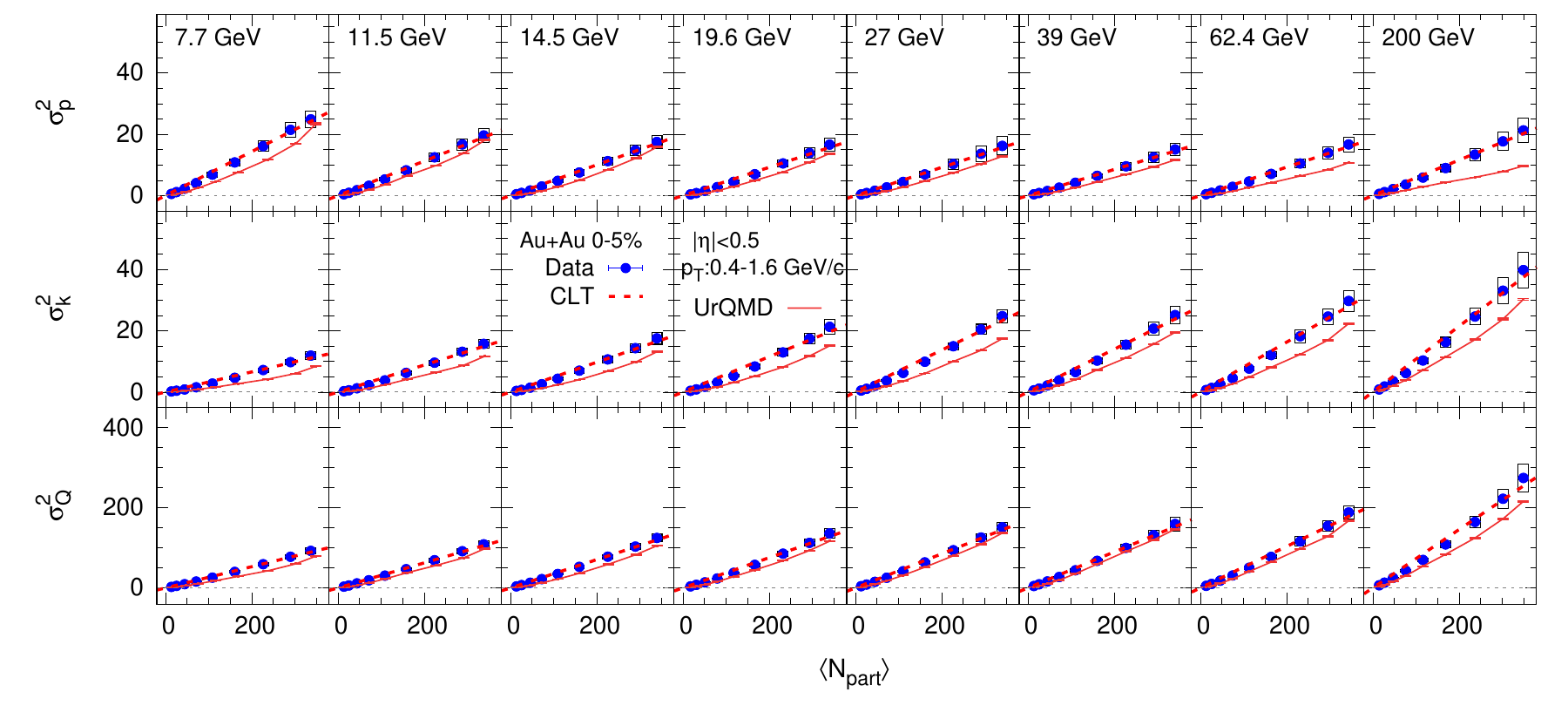}
		\caption{Centrality dependence of efficiency-corrected second-order diagonal cumulants (variances) of net-proton, net-kaon and net-charge (top to bottom) of the multiplicity distributions for \gold collisions at \sNN~= 7.7, 11.5, 14.5, 19.6, 27, 39, 62.4 and 200 GeV (left to right) within kinematic range $|\eta|<0.5$ and $0.4<p_{\text{T}}<1.6$ GeV/\textit{c}. The boxes represent the systematic error. The statistical error bars are within the marker size. The dashed lines represent scaling predicted by central limit theorem and the solid lines are UrQMD calculations.}
		\label{variance}
	\end{figure*}

	\begin{figure*}[th]
		\includegraphics[width=\textwidth]{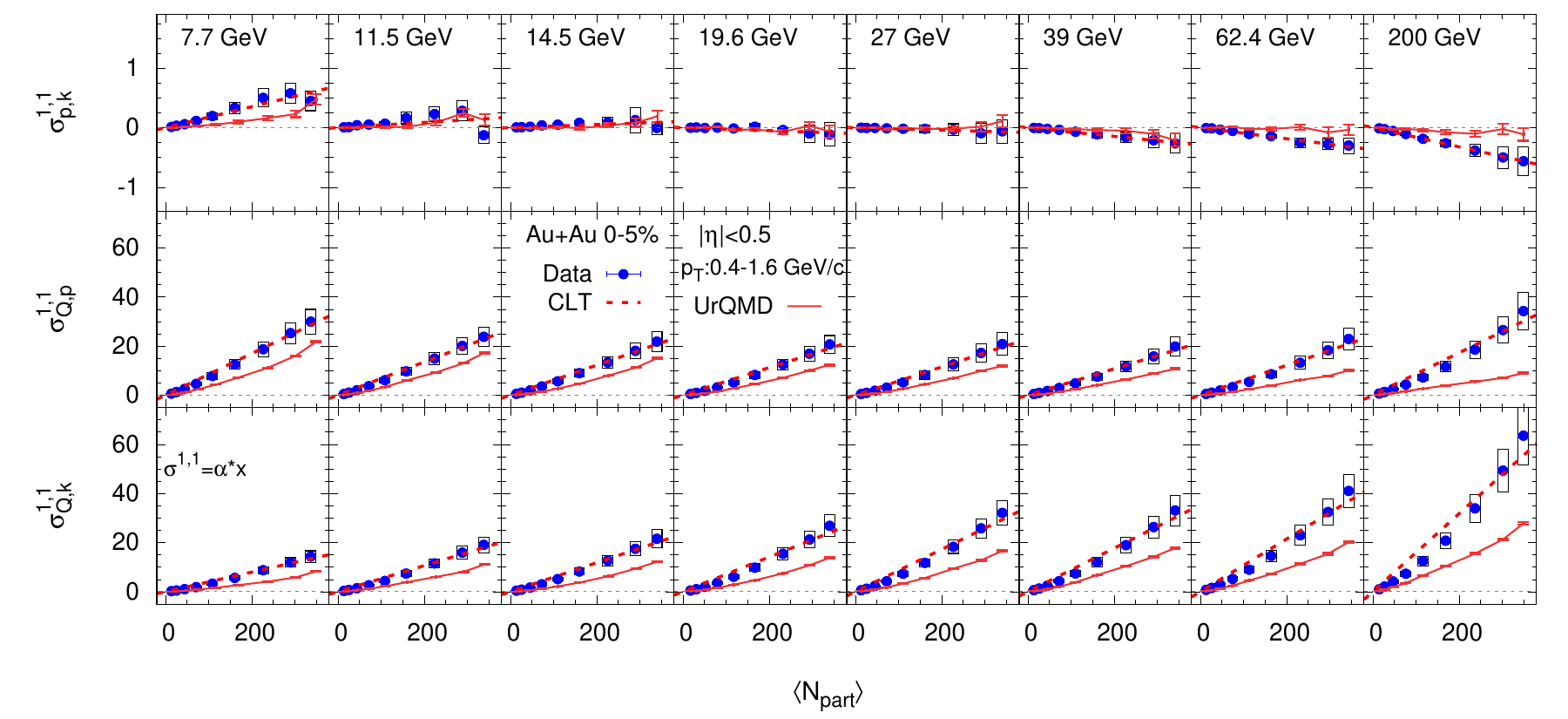}
		\caption{Centrality dependence of second-order off-diagonal cumulants of net-proton, net-charge and net-kaon for \gold collisions at \sNN~= 7.7, 11.5, 14.5, 19.6, 27, 39, 62.4 and 200 GeV (left to right) within kinematic range $|\eta|<0.5$ and $0.4<p_{\text{T}}<1.6$ GeV/\textit{c}. Error bars are statistical and boxes are systematic errors. The dashed lines represent scaling predicted by the central limit theorem and the solid lines are UrQMD calculations.}
		\label{covariance}
	\end{figure*}
	%


	\section{Results and discussion}
	
	\label{sce:results}
	%
	We start with the differential measurements of the cumulants. Figure~\ref{deltaeta} shows the efficiency-corrected diagonal and off-diagonal cumulants as a function of the $\eta$-window for most central (0-5$\%$) and peripheral (70-80$\%$) bins and for eight different collision energies. Since our measurements involve centrality determination using charged tracks within the acceptance of $0.5<|\eta|<1$, we can vary the width of the $\eta$-window to a maximum value of 0.5. 
	We observe that in central events the cumulants, except \SigPK~show a linear increasing trend with increasing $\eta$-window within the range of $0.1<|\eta|<0.5$ for the measured beam energies. \SigPK~shows significantly different trends in contrast to the other cumulants. It is negative at all energies except for \sNN~= 7.7 GeV.
	As discussed below, this might indicate an anti-correlation between proton and kaon production, as expected from the fact that positive baryon number is associated with negative strangeness~\cite{Koch:2005vg}. At the lowest beam energy, other mechanisms~\cite{Balewski:1998pd,Zhou:2017jfk}~must dominate such anti-correlation to change the sign of \SigPK. 
	The magnitudes of all the cumulants are closer to zero at $|\eta|<0.1$; 
	for peripheral collisions (70-80\%) the cumulants are close to zero over the whole range of $\eta$-window.
	Ref.~\cite{Ling:2015yau} discusses the underlying origin of the rapidity acceptance ($\Delta y_{\rm window}$) dependence of cumulants. The authors argue that a linear dependence ($\kappa^\alpha \propto \Delta y_{\rm window}$) is expected if the cumulants are driven by uncorrelated contributions developed over a range of acceptance ($\Delta y_{\rm corr}$) that is much smaller than the window of measurements ($\Delta y_{\rm window}$). If the underlying correlations are developed over a range $\Delta y_{\rm corr}\gg\Delta y_{\rm window}$, one expects deviations from a linear dependence. Although we use pseudorapidity rather than rapidity, based on the motivations from~\cite{Ling:2015yau}, we perform linear fits ($a+b\times |\eta|$) to the data shown in Fig.~\ref{deltaeta} for 0-5\%. Similar linear growth is also observed for 70-80\% centrality. We do not find a significant deviation from linear dependence within the range of our measurements. 
	However, it is known that such linear growth will saturate at a certain $\eta$-window, and then should decrease to a minimum value at $|\eta| = 2 Y_{\rm beam}$ due to the global charge conservation~\cite{Koch:2005vg}. A detailed simulation demonstrating the effect of global conservation, using the UrQMD and HRG models, can be found in~\cite{Chatterjee:2016mve}. Figure~\ref{deltaeta} shows UrQMD calculations for $0-5\%$ centrality. UrQMD explains the diagonal cumulants but does a poor job for the off-diagonal ones. This already hints that off-diagonal cumulants contain additional information as compared to diagonal cumulants and cannot be described by hadronic models. 

	It will be possible to perform an improved study of the acceptance dependence of cumulants with the future iTPC upgrade of STAR planned for the BES-II program at RHIC~\cite{STARnote:598, STARnote:644}. For the BES-II program, the centrality determination can be performed by an independent event plane detector (EPD)~\cite{STARnote:666} over an acceptance window of $2.1<\eta<5.1$. Therefore, it will be possible to measure acceptance dependence of the cumulants using iTPC over a wider $\eta$-window ($\sim1.7$) and search for deviations from a linear trend as predicted in~\cite{Koch:2005vg, Ling:2015yau, Brewer:2018abr}. 
	\begin{figure*}[t]
		\includegraphics[width=\textwidth]{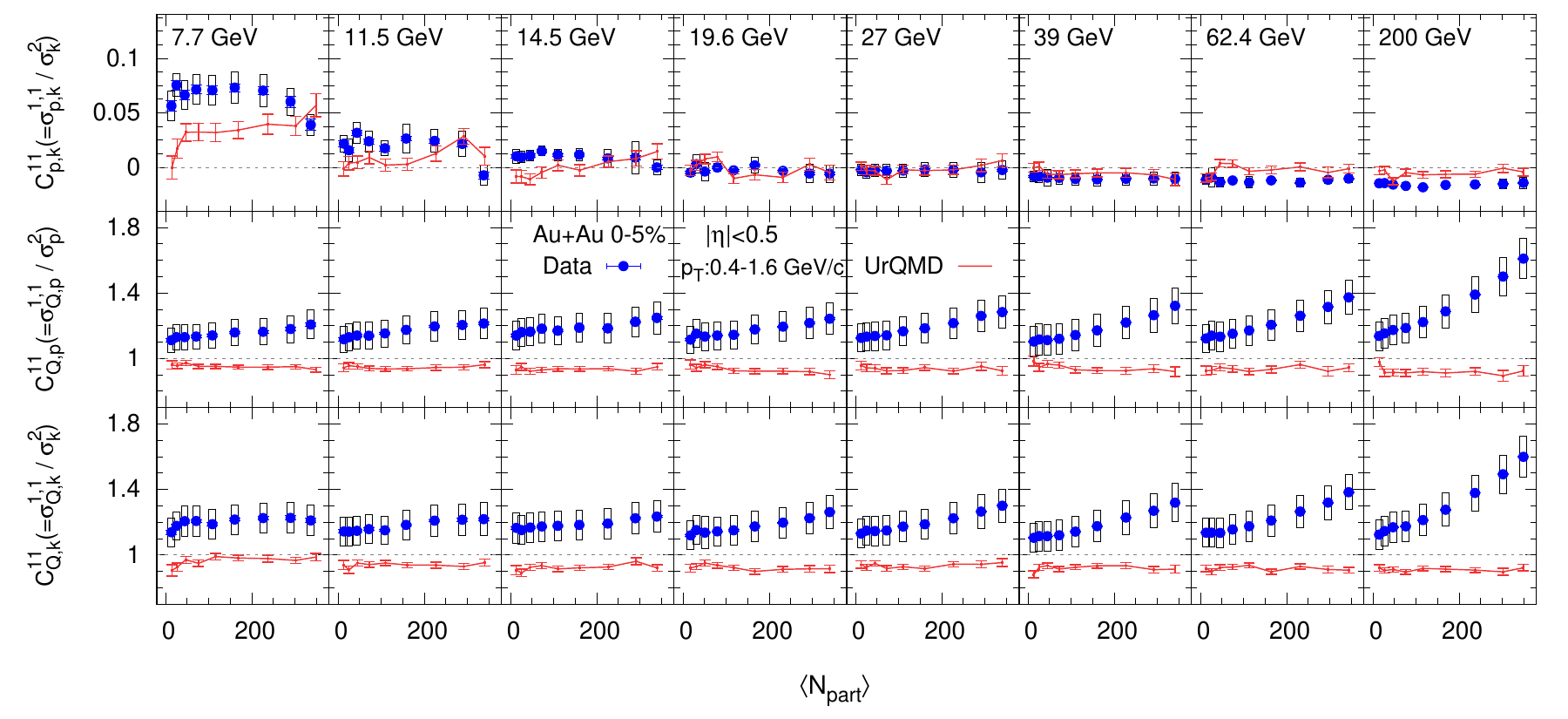}
		\caption{Centrality dependence of second-order off-diagonal to diagonal cumulants ratios of net-proton, net-charge and net-kaon for Au+Au collisions at \sNN~= 7.7, 11.5, 14.5, 19.6, 27, 39, 62.4 and 200 GeV (left to right) within the kinematic range $|\eta|<0.5$ and $0.4<p_{\text{T}}<1.6$ GeV/\textit{c}. Error bars are statistical and boxes are systematic errors. The solid lines represent the UrQMD calculations.}
		\label{ratio} 
	\end{figure*}

	For the rest of the paper, we present results for cumulants integrated over the window of $|\eta|<0.5$. In Figs.~\ref{variance} and~\ref{covariance} we present the centrality dependence of efficiency-corrected second-order diagonal and off-diagonal cumulants, respectively, for all eight energies. 
	For all diagonal cumulants shown in Fig.~\ref{variance}, we find a linear increasing trend as expected from a scaling predicted by the central limit theorem (CLT): $\sigma^{2} ~\propto$~\Npart. The slopes of $\sigma^{2}_k$ and $\sigma^{2}_Q$ show a monotonic increase with the collision energy. A different trend is seen for the \Npart~dependence of $\sigma^{2}_p$ for net-proton distributions. The slope of this dependence decreases in the range of \sNN~= 7.7-19.6 GeV, remains approximately constant over \sNN~= 19.6-39 GeV and then increases in the range of \sNN~= 39-200 GeV. 
	Such a trend, first reported in ~\cite{Adamczyk:2013dal}, can be attributed to the details of baryon transport that has a strong collision energy dependence. 
	As can be seen from Fig.~\ref{variance}, at \sNN~= 200 GeV, $\sigma^{2}_{Q} > \sigma^{2}_{k} > \sigma^{2}_{p}$, while for the range of \sNN~= 7.7-19.6 GeV, one finds an ordering like $\sigma^{2}_{Q} > \sigma^{2}_{p} > \sigma^{2}_{k}$ as expected from a baryon dominated medium at lower energies. We find that UrQMD calculations slightly underestimate these cumulants although seem to qualitatively describe the trend seen in data. 
	
	The centrality dependence of the off-diagonal cumulants \SigQK and \SigQP, shown in Fig.~\ref{covariance}, is very similar to that of the diagonal cumulants. A distinct difference is seen for \SigPK. The values of \SigPK~are negative at higher energies. At lower energies, we observe a slight deviation from CLT associated with a sign change that we discussed previously in the context of Fig.~\ref{deltaeta}. The magnitude of \SigPK~is much smaller than \SigQK~(or \SigQP) as the latter can have a contribution from self-correlations. Once again we see quantitative disagreement between data and UrQMD calculations which is more pronounced in comparison with what is seen for the diagonal cumulants. 
	
	We now explore the order of magnitude difference between \SigPK~and \SigQP~(or \SigQK) by constructing ratios of off-diagonal and diagonal cumulants defined as
	%
	\bea
	C_{p,k} = \frac{\sigma^{1,1}_{p,k}}{\sigma^{2}_{k}},
	C_{Q,k} = \frac{\sigma^{1,1}_{Q,k}}{\sigma^{2}_{k}},
	C_{Q,p} = \frac{\sigma^{1,1}_{Q,p}}{\sigma^{2}_{p}}.
	\label{eq_koch}
	\eea
	The construction of $C_{\alpha,\beta}$, also referred to as ``Koch ratio", is motivated by~\cite{Koch:2005vg}. The trivial volume dependence of the cumulants is expected to be cancelled in such ratios. Also, since the number of $p(\bar{p})$ and $K^\pm$ are subsets of $Q^\pm$, it is natural to normalize $\sigma^{1,1}_{Q,k}$ ($\sigma^{1,1}_{Q,p}$) by the self-correlation of net-kaon (net-proton). It must be noted that $\sigma^{1,1}_{p,k}$ is not affected by trivial self correlations. One can, therefore, choose either ${\sigma^{2}_{k}}$ or ${\sigma^{2}_{p}}$ in the denominator of $C_{p,k}$; in this paper we use $\sigma^{2}_{k}$. Note that, in the original definition of $C_{B,S} = -3\times \chi^{1,1}_{B,S}/\chi^{2}_{S}$, the authors of \cite{Koch:2005vg} included a pre-factor of $-3$; for our definition in Eq.~\ref{eq_koch} we do not include such pre-factors. 
	
	Figure~\ref{ratio} presents the centrality dependence of these Koch ratios. An interesting trend is seen for $C_{p,k}$. It shows a weak centrality dependence and a sign change as expected from the trend observed for \SigPK~in Fig.~\ref{covariance}. For most of the centrality bins, the sign change happens around 14.5-19.6 GeV. We will come back to this important observation later in this paper. 
	On the other hand, $C_{Q,k}$ and $C_{Q,p}$ show much stronger energy and centrality (particularly, at higher energies) dependence. Since they measure the excess correlation, it is not obvious why an increase of net-charge is strongly affected by the increase of net-proton or net-kaon in the system. We see both qualitative and quantitative disagreements between data and UrQMD calculations. We investigate this in the following sub-section by concentrating only on two centrality bins.

	\begin{figure}[t]
		\includegraphics[width=0.48\textwidth]{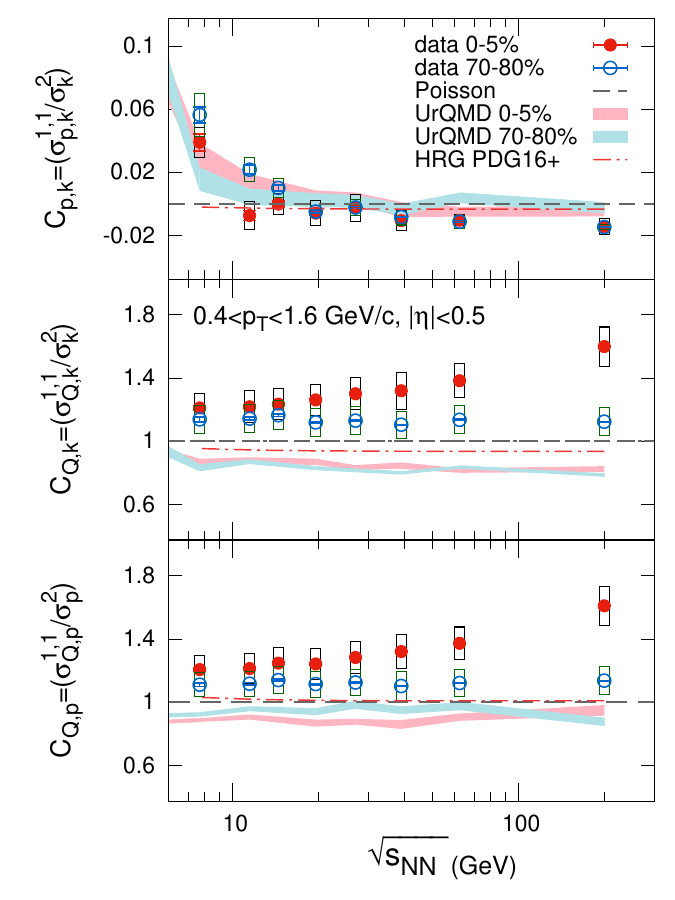}
		\caption{Beam energy dependence of cumulant ratios ($C_{p,k}, C_{Q,k}$ and $C_{Q,p}$; top to bottom) of net-proton, net-kaon and net-charge for \gold collisions at \sNN~= 7.7, 11.5, 14.5, 19.6, 27, 39, 62.4 and 200 GeV. The bands denote the UrQMD calculations for 0-5\% and 70-80\% central collisions and the HRG values are denoted by dotted lines. Poisson baseline is denoted by dotted lines. Error bars are statistical and boxes are systematic errors.}
		\label{energy} 
	\end{figure}

	Figure~\ref{energy} shows the beam energy dependence of the $C_{p,k}, C_{Q,k}$ and $C_{Q,p}$ for two centralities (0-5\% and 70-80\%). We compare the data with the UrQMD~\cite{Bass:1998ca} calculations and with an implementation of the HRG model based on the experimentally known hadron spectrum (PDG)~\cite{Bellwied:2018tkc}.  
	%
	%
	%
	
	Correlated fluctuations of total kaons and protons were previously reported by NA49 and STAR collaborations in~\cite{Anticic:2011am, Abdelwahab:2014yha}. However, in this work, we measure the correlation in the corresponding net-multiplicity distributions to study net-baryon and net-strangeness correlations in a more direct way. The top panel of Fig.~\ref{energy} indicates that $C_{p,k}$ has a very weak energy dependence down to 19.6 GeV that is very similar for both the central and peripheral events. 
	The UrQMD model seems to give rise to a $C_{p,k}$ that is either positive or consistent with zero within the uncertainties. 
	%
	On the other hand, the HRG model calculations for $C_{p,k}$ are consistent with zero. Clearly, we do not see such trends in the data. For the two centralities shown in Fig.~\ref{energy} (and for all the centralities shown in Fig.~\ref{ratio}) we see that $C_{p,k}$ is significantly negative (3$\sigma$ below zero at \sNN~= 200 GeV) at higher energies. At lower energies, $C_{p,k}$ becomes positive (4$\sigma$ above zero at \sNN~= 7.7 GeV). 
	The contribution to $C_{p,k}$ from a hadronic medium is difficult to understand. The decay of resonance $\Lambda(1520)\rightarrow p + K^-$ with a branching ratio of ($22.5 \pm 0.5\%$) ~\cite{Patrignani:2016xqp} can contribute to $C_{p,k}$. However, such a decay increases net-proton and decreases net-kaon in the system and therefore, can only lead to an anti-correlation and cannot be responsible for the positive values of $C_{p,k}$ at lower energies. 
	An indirect source of correlations between net-proton and net-kaon is expected to arise at lower energies from the associated production: $pp\rightarrow p \Lambda(1115) K^+$~\cite{Balewski:1998pd}. Such a hadronic scattering process dominates owing to the abundance of protons and leads to an increase in the fraction of net-kaon (and also net-lambda) at lower energies~\cite{Andronic:2005yp,Adamczyk:2017wsl}. One, therefore, expects events with higher net-protons to be associated with higher net-kaons resulting in positive values of $C_{p,k}$ at lower energies. The associated production is already included in the UrQMD model~\cite{Zhou:2017jfk}, which might explain the trend seen in Fig.~\ref{energy}. Note that the associated production is followed by the resonance decay $\Lambda(1115)\rightarrow p + \pi^-$ with a branching ratio of $63.9 \%$. Since the decay proton from this channel is strongly correlated with the $K^+$ from the associated production, one expects a further increase in the net-proton to net-kaon correlation as energy decreases. The UrQMD calculations shown in Fig.~\ref{energy} correspond to an evolution time of $\tau_{\text{evol}}=100$ fm/$c$ and do not include the decay of $\Lambda(1115)$ that has a decay length of $c\tau=7.89$ cm. Although we apply a DCA cut of 1 cm in our analysis we do not fully exclude the protons coming from the $\Lambda(1115)$ decays. Therefore, we force the decay of all the produced $\Lambda$'s in UrQMD and find an increase of $C_{p,k}$ by about $30\%$ at 7.7 GeV. At higher energies (200 GeV) we find negligible effect on $C_{p,k}$ from both associated production and the subsequent $\Lambda(1115)$ decay.
	%
	%
	%
	At higher energies where $\mu_B$ is small, the abundance of baryonic resonances like $\Lambda(1520)$ is also small~\cite{Adams:2006yu, ALICE:2018ewo}. This may be the possible reason for nearly zero values of $C_{p,k}$ seen in HRG and for UrQMD at higher energies. 
	%
	%
	Therefore, the negative value of $C_{p,k}$ at higher energies may not be dominantly coming from the hadronic phase. We discuss the expectations from a QGP phase below.

	The correlated production of net-proton and net-kaons from a QGP phase is a consequence of 
	%
	%
	positive strangeness (carried by a strange anti-quark) being associated with a negative baryon number. One, therefore, expects production of net-strangeness or net-kaon to be correlated with a compensating decrease in net-baryon or net-proton. 
	This strong anti-correlation between net-strangeness and net-baryon in the QGP phase is expected to have weak $T$ and $\mu_B$ dependence~\cite{Koch:2005vg}. In a hadronic phase, such correlations will have a strong dependence on $T$ and $\mu_B$. One of the original predictions of~\cite{Koch:2005vg} was that $C_{B, S}$ would show weak $T$ and $\mu_B$ dependence in the QGP phase and a strong dependence in the HRG phase. Since changing \sNN~changes both $T$ and $\mu_B$, it is not straightforward to directly compare the \sNN~dependence of $C_{p,k}$ shown in Fig.~\ref{energy} to the behavior as predicted for $C_{B,S}$ in ~\cite{Koch:2005vg}. 
	%
	%
	%
	%
	%
	%
	%
	%
	%
	%
	Nevertheless, the current data on $C_{p,k}$ may provide some important insights on the baryon-strangeness correlations that are expected to change at the onset of deconfinement~\cite{Koch:2005vg,Bazavov:2014xya}. 
	
	A very different behavior is observed for the energy dependence of $C_{Q,p}$ and $C_{Q,k}$. Both of these ratios show significantly higher correlations in central 0-5\% events than in 70-80\% events. The difference shows an increasing trend with energy, not predicted by UrQMD calculations. The HRG predictions for these ratios are much lower than the data. Clearly, the excess correlation of net-charge with net-kaon and net-proton, cannot be explained by either thermal (HRG) or non-thermal (UrQMD) production of hadrons. It must be noted that unlike $C_{p,k}$ one expects many resonances to contribute to $C_{Q,p}$ and $C_{Q,k}$. For example, in case of $C_{Q,p}$, one expects contributions from the decay of baryons such as $\Delta^{++}\rightarrow \pi^+ + p$~\cite{Patrignani:2016xqp}. The doubly charged state of $\Delta^{++}$ can simultaneously increase net-proton and net-charge. 
	%
	%
	%
	The quantity $C_{Q,k}$ should have contributions from the resonance decay to $K$ and $\pi$ states that has a net-charge state and can decay to change the number of net-kaons. Resonance decays like $K^{*0}(892) \rightarrow\pi^\pm+K^\mp$ or $\phi (1020) \rightarrow K^+ + K^- $ ~\cite{Patrignani:2016xqp} will not change $C_{Q,k}$ as they do not lead to correlated production of net-charge and net-kaon. Decays like $K^{*\pm}(892) \rightarrow K_{S}^{0} + \pi^\pm$ increases both the net-strangeness and net-charge in the system, although, it is not clear if such decays lead to correlated production of net-kaon and net-charge. 
	Therefore, a small contribution to $C_{Q,k}$ from the hadronic phase is expected. More theoretical input is needed to see if the excess correlations, seen for $C_{Q,p}$ and $C_{Q,k}$, indeed come from the resonance states that have not been included in the existing hadronic models~\cite{Chatterjee:2017yhp}. 
	It will also be important to understand if the growth of these cumulants with collision energy can be explained by model calculations that include contributions from the QGP phase.

	\section{Summary and outlook}\label{sec:summary}
	In this paper we present the second-order diagonal and off-diagonal cumulants of net-charge, net-proton and net-kaon multiplicity distributions, within a common acceptance of $|\eta|<0.5$ and $0.4<p_{\text{T}}<1.6$ GeV/\textit{c} in Au+Au collisions at eight different energies in the range of \sNN~= 7.7-200 GeV. The primary motivation of this analysis is to understand the mechanism behind the correlated production of hadrons carrying different conserved charges in heavy-ion collisions. Many theoretical calculations hint that correlated production of two different conserved charges contains additional information that can provide crucial tests for hadronic models of heavy-ion collisions. 
	%
	%
	With the current measurements we indeed demonstrate that although hadronic models describe the variance of a particular conserved charge distribution, they fail to describe many features of the correlated fluctuations of two different kinds of conserved charges. 

	The findings of this analysis can be summarized as follows. We observe a strong dependence of the cumulants with the phase space window of measurements. When plotted as a function of the $\eta$-window, all cumulants show an approximately linear dependence, a trend that is reproduced by UrQMD model calculations, although, the growth of the off-diagonal cumulants is weaker in UrQMD than in data. The centrality dependence of the cumulants within a given pseudorapidity window ($|\eta| < 0.5$) is also linear when plotted against the number of participants. The slope of such dependence for the \SigPK~changes sign at lower energies. We construct the Koch ratios $C_{p,k}, C_{Q,p}$ and $C_{Q,k}$ by dividing the off-diagonal cumulants by the diagonal ones to remove the trivial volume dependence. The values of $C_{p,k}$ are clearly negative (with about 3$\sigma$ significance) at \sNN~= 200 GeV, they change sign around 19.6 GeV for most centrality bins, and become positive (with about 4$\sigma$ significance) at \sNN~= 7.7 GeV. UrQMD and HRG predict values of $C_{p,k}$ that are either positive or consistent with zero and do not explain the non-zero negative values observed for data at higher energies. We argue that the energy and centrality dependence of $C_{p,k}$ will help understand the baryon-strangeness correlations that is predicted to have different dependence on $T$ and $\mu_B$ between the QGP and hadronic phases~\cite{Koch:2005vg,Majumder:2006nq,Bazavov:2013dta,Bazavov:2014xya}. 
	%
	%
	
	The ratios $C_{Q,p}=\sigma^{1,1}_{Q,p}/\sigma^2_{p}$ and $C_{Q,k}=\sigma^{1,1}_{Q,k}/\sigma^2_{k}$ are constructed such that they measure the excess correlations of net-charge with net-proton and net-kaon, respectively. This removes the trivial self-correlations arising from the fact that $Q^\pm$ contains both $p(\bar{p})$ and $K^\pm$. Both $C_{Q,p}$ and $C_{Q,k}$ show strong centrality dependence in data indicating the presence of a large excess correlation in most central events in comparison with peripheral events. The difference between central and peripheral events seems to grow with energy. Both UrQMD and HRG models under-predict the data and can not describe the strong energy and centrality dependence of $C_{Q,p}$ and $C_{Q,k}$. Current data will, therefore, constrain HRG and improve modeling of correlated production of particles carrying different conserved charges in heavy-ion collisions. It will be important to obtain theoretical input to see if the behavior of $C_{Q,p}$ and $C_{Q,k}$ has a partonic origin and therefore is not captured by conventional hadronic models. Finally, we argue that the measurements of the full cumulant matrix of net-multiplicity distributions in a common acceptance will improve the estimation of freeze-out parameters extracted by HRG or lattice calculations that help map the QCD phase diagram.

	%
	
	The measurements presented here are limited by the current acceptance of the STAR detector. 
	%
	A more comprehensive measurement of higher-order cumulants will be pursued by the second phase of BES program (BES-II) with better capability of centrality determination using the EPD and with the improved acceptance of the inner Time Projection Chamber (iTPC) upgrade of STAR. Also, in this paper we have restricted ourselves to the measurements of off-diagonal cumulants up to second-order. With higher-statistics data sets and improved techniques of detector efficiency corrections it will be possible to measure higher-order off-diagonal cumulants in the upcoming BES-II program of RHIC. \\
	\section{ACKNOWLEDGMENTS}
	
	We are grateful to Jacquelyn Noronha-Hostler for providing the HRG model calculations. We are thankful to Jorge Noronha, Sandeep Chatterjee, Sayantan Sharma, Swagato Mukherjee, Frithjof Karsch, Volodymyr Vovchenko, Sourendu Gupta, Rajiv V. Gavai and Che Ming Ko for the fruitful discussions.  
	We thank the RHIC Operations Group and RCF at BNL, the NERSC Center at LBNL, and the Open Science Grid consortium for providing resources and support.  This work was supported in part by the Office of Nuclear Physics within the U.S. DOE Office of Science, the U.S. National Science Foundation, the Ministry of Education and Science of the Russian Federation, National 
	Natural Science Foundation of China, Chinese Academy of Science, the Ministry of Science and Technology of China and the Chinese Ministry of Education, the National Research Foundation of Korea, Czech Science Foundation and Ministry of Education, Youth and Sports of the Czech Republic, Department of Atomic Energy and Department of Science and Technology of the Government of India, the National Science Centre of Poland, the Ministry  of Science, Education and Sports of the Republic of Croatia, RosAtom of Russia and German Bundesministerium fur Bildung, Wissenschaft, Forschung and Technologie (BMBF) and the Helmholtz Association.
	
	\bibliographystyle{apsrev4-1}
	\bibliography{offdiagonal}
	




\end{document}